\documentclass[fleqn,usenatbib,useAMS]{mnras}
\pdfoutput=1 
\usepackage{amsmath,amstext}
\usepackage{newtxtext,newtxmath}
\usepackage[T1]{fontenc}
\usepackage{ae,aecompl}
\usepackage{graphicx}
\usepackage[figure,figure*]{hypcap}
\usepackage{url}
\usepackage{hyperref}
\usepackage{natbib}
\usepackage{comment}

\newcommand\phn{\phantom{0}}

\title[Kepler DR25 M Star Planet Occurrence Rates]{Occurrence Rates of Planets Orbiting M Stars: Applying ABC to \emph{Kepler} DR25, \emph{Gaia} DR2, and 2MASS Data}

\author[Hsu, Ford and Terrien]{Danley C. Hsu,$^{1,2,3,4}$
Eric B. Ford$^{1,2,3,4}$
and Ryan Terrien$^{5}$
\\
$^{1}$Department of Astronomy \& Astrophysics, 525 Davey Laboratory, The Pennsylvania State University, University Park, PA, 16802, USA\\
$^{2}$Center for Exoplanets and Habitable Worlds, 525 Davey Laboratory, The Pennsylvania State University, University Park, PA, 16802, USA\\
$^{3}$Center for Astrostatistics, 525 Davey Laboratory, The Pennsylvania State University, University Park, PA, 16802, USA\\
$^{4}$Institute for Computational \& Data Sciences, The Pennsylvania State University, University Park, PA, 16802, USA\\
$^{5}$Department of Physics and Astronomy, Carleton College, Northfield, MN 55057, USA}

\date{Accepted XXX. Received YYY; in original form ZZZ}

\pubyear{2020}

\begin{document}
\label{firstpage}
\pagerange{\pageref{firstpage}--\pageref{lastpage}}
\maketitle

\begin{abstract}
We present robust planet occurrence rates for \emph{Kepler} planet candidates around M stars for planet radii $R_p = 0.5-4~\textrm{R}_\oplus$ and orbital periods $P = 0.5-256$ days using the approximate Bayesian computation (ABC) technique.  
This work incorporates the final \emph{Kepler} DR25 planet candidate catalog and data products and augments them with updated stellar properties using \emph{Gaia} DR2 and 2MASS PSC. 
We apply a set of selection criteria to select a sample of 1,746 \emph{Kepler} M dwarf targets that host 89 associated planet candidates. 
These early M dwarfs and late K dwarfs were selected from cross-referenced targets using several photometric quality flags from \emph{Gaia} DR2 and color-magnitude cuts using 2MASS magnitudes.  
We estimate a habitable zone occurrence rate of
$f_{\textrm{M,HZ}} = 0.33^{+0.10}_{-0.12}$
for planets with $0.75-1.5$ R$_\oplus$ size.
We caution that occurrence rate estimates for \emph{Kepler} M stars are sensitive to the choice of prior due to the small sample of target stars and planet candidates.  
For example, we find an occurrence rate of  $4.2^{+0.6}_{-0.6}$ or $8.4^{+1.2}_{-1.1}$
planets per M dwarf (integrating over $R_p = 0.5-4~\textrm{R}_\oplus$ and $P = 0.5-256$ days) for our two choices of prior.  
These occurrence rates are greater than those for FGK dwarf target when compared at the same range of orbital periods, but similar to occurrence rates when computed as a function of equivalent stellar insolation.  
Combining our result with recent studies of exoplanet architectures indicates that most, and potentially all, early-M dwarfs harbor planetary systems.  
\end{abstract}

\begin{keywords}
planets and satellites: fundamental parameters -- planets and satellites: detection -- methods: statistical -- 
planetary systems -- stars: statistics -- stars: low-mass
\end{keywords}

\section{Introduction}
\label{secIntro}
M dwarf stars represent an interesting stellar type for occurrence rate studies.  
Transiting exoplanets are more easily detectable around M stars than sun-like stars because transit depth $\delta \propto R_*^{-2}$ (where $R_*$ is the host star's radius) and M stars are smaller than FGK stars.  The habitable zone (HZ) around M stars is also closer to the star due to their lower luminosity.  Therefore, HZ planets around M stars have smaller periods in comparison to HZ planets around FGK stars, increasing both their geometric transit probability and their transit SNR for a transit survey with fixed lifetime.  Both these considerations make small planets around M stars relatively favorable targets for 
follow-up transit observations.

Contrasting the M dwarf exoplanet population to the exoplanet population around FGK dwarfs can provide significant insights to planet formation theories.  Since the stellar types have different accretion and disk dissipation timescales, observing clear differences in their exoplanet populations can give insight into the relative importance of the two timescales \citep{EP2017}.
M dwarfs also have significantly less luminosity compared to FGK dwarfs, so differences in the exoplanet populations could point to the relative significance of stellar irradiance and disk properties for planetary accretion and migration. 
For these reasons, many projects have and/or will search for exoplanets orbiting M dwarfs, including the ground-based MEarth \citep{IBC+2015} and SPECULOOS \citep{DGQ+2018} transit surveys focusing on M dwarfs.  Additionally, space-based surveys by the \emph{K2} \citep{HSH+2014} and \emph{TESS} \citep{RWV+2015} missions have significant increased the number of nearby M dwarfs surveyed for transiting planets with small orbital periods. 
Ground-based radial velocity (RV) surveys using instruments such as HPF \citep{MRB+2012}, SPIRou \citep{AKD+2014}, and CARMENES \citep{QAR+2018} also will target M dwarfs to better characterize their exoplanet population, extending to larger orbital periods.

Several studies have attempted to quantify occurrence rates for M stars using \emph{Kepler} data.  An early study by \citet{DC2013} used the Inverse Detection Efficiency Method (IDEM), which assigns a completeness ``weight'' to each detection
to estimate the number of \emph{Kepler} targets that needed to be observed to achieve that one detection. That study determined the occurrence rate of planets with radii $<4 R_\oplus$ for stars believed to be cooler than $4000$ K from Q1-Q6 \emph{Kepler} data.  \citet{KRK+2013,G2013} used similar methods, but arrived at larger estimates for the HZ rate, primarily due to different HZ limits.  \citet{MS2014} took a slightly different approach and modeled each planet using a weighted kernel density estimator, arriving at the conclusion that previous rates were underestimated by a factor of 1.6 due to incompleteness.  \citet{GML+2014} made use of an iterative simulation model to characterize the dependence of occurrence rate on planet radius, concluding that the distribution of rates peaks at planet radius $R_p \sim 0.8~\textrm{R}_\oplus$ and that on average M dwarfs host approximately 2 planets with planet radii $R_p = 0.5-6~\textrm{R}_\oplus$ and orbital period $P < 180$ days.

A particularly influential study is that of \citet{DC2015}, which characterized occurrence rates for cool stars with $T_\textrm{eff} < 4000$ K and $\log g > 3$ using the Q1-Q17 DR24 \emph{Kepler} data and adding archival spectroscopic and photometric observations to improve their estimates of stellar parameters.  
Using a detection efficiency curve that accounts for their custom pipeline's completeness as characterized by analyzing synthetic data with injected planets and applying a correction for contamination due to false positives, they produced smoothed maps of planet detection completeness to arrive at binned occurrence rates for $P = 0.5-200$ days and $R_p = 0.5 - 4~\textrm{R}_\oplus$.  They conclude that on average there are $2.5\pm0.2$ planets with $R_p = 1-4~\textrm{R}_\oplus$ and $P < 200$ days. 
A study by \citet{GMK+2016} also using the DR24 \emph{Kepler} data found occurrence rates of $f = 2.2\pm0.3$ over $R_p = 1-4~\textrm{R}_\oplus$ and $P = 1.2-180$ days, consistent with those of \citet{DC2015}.   
A study by \citet{MPA2015} which selected M dwarfs having $T_\textrm{eff} = 2400-3865$ K found systematically lower estimated occurrence rates than \citet{DC2015} over $R_p = 0.5-4~\textrm{R}_\oplus$ and $P = 0.5-50$ days. \citet{MPA2015} suggests this difference is likely the result of both studies applying the \citet{DC2015} detection efficiency model, but having used different planet detection pipelines.

M stars span a large range of stellar masses and temperatures.  
Given \emph{Kepler}'s aperture and target list, most occurrence rate studies of M dwarfs using \emph{Kepler} data have been based on early-type M dwarfs and likely had significant contamination from late K dwarfs due to difficulties in accurately dividing K dwarfs and M dwarfs using stellar properties derived from photometry prior to Gaia DR2. 
While \emph{Kepler} only observed a small number of mid-type M dwarfs, very few studies have focused on occurrence rates for this challenging sample.
The latest study to focus on mid-type M dwarfs is \citet{HCM+2019}, which used \emph{Kepler} DR25 data and enhanced it with \emph{Gaia} DR2 \citep{GBV+2018} and spectroscopic measurements to derive more accurate stellar properties.  
They found a significantly enhanced occurrence rate for planets with $R_p = 0.5-2.5~\textrm{R}_\oplus$ and $P < 10$ days compared to previous studies which computed occurrence rates based on samples dominated by of earlier stars (e.g., \citet{DC2015} and \citet{MPA2015}).  
\citet{HCM+2019} also concluded that occurrence rates decreased as a function of stellar type.

A small number of studies have begun to characterize the M dwarf exoplanet population via their planetary system architectures.
\citet{MMV+2015} calculated the occurrence rate of compact multiple systems (i.e., systems with multiple planets at $P < 10$ days) for mid-type M stars using DR24 \emph{Kepler} data, concluding that $21^{+7}_{-5}\%$ of mid-M dwarfs host such systems.  
A separate study by \citet{BJ2016} investigated planet multiplicity for a sample of primarily early M dwarfs and concluded that the \emph{Kepler} dichotomy also exists in the exoplanet population around M dwarf.
The architectures inferred for M dwarf hosts can be compared to the architectures of FGK host stars \citep[e.g.,][]{MPA+2018,HeFR2019}.

This study aims to provide the most up-to-date occurrence rate estimates using the final \emph{Kepler} DR25 planet catalog and associated data products and an enhanced set of stellar properties using \emph{Gaia} DR2 measurements.  
In \S\ref{secModel} we briefly review how we adapt our previous methodology from \citet{HFR+2019} to this study.  
In \S\ref{secResults} we cover the selection of our M dwarf catalog using \emph{Kepler} DR25, \emph{Gaia} DR2, and 2MASS  2MASS Point Source Catalog (PSC) measurements as well as present our new occurrence rate estimates.  Lastly, in \S\ref{secDiscussion} we discuss the implications of our results, focusing on comparisons with our previously estimated FGK rates and other M dwarf occurrence rate studies.

\section{Methodology}
\label{secModel}
We apply a Bayesian hierarchical model to perform inference using a flexible model for the occurrence rate of planets per star as a function of size and orbital period.  
We present results using two parameterizations (and associated priors) for the occurrence rates (detailed in \S\ref{secImproveParam}) to help assess the sensitivity to our choice of priors. 
While our population model is simple, we employ a sophisticated model for the detection and vetting efficiency of the \emph{Kepler} mission and pipeline.  
The complexity of the \emph{Kepler} model makes it impractical to write down a likelihood analytically.
Therefore, we make use of Approximate Bayesian Computation (ABC) which is designed to perform Bayesian inference for a real world problems when writing a rigorous likelihood is intractable or impractical.  
Our study follows most of the same methodology that was described and validated in \citet{HFR+2019}.  Our ABC implementation \citep{ABC_Julia} and Exoplanets Systems Simulator (SysSim) forward model \citep{ExoplanetsSysSim2020} are publicly available on Github\footnote{The cited releases used in this study implement an older version (v0.6) of the Julia programming language. For code that is compatible with the current release of Julia (v1.4), please use the code available at \url{https://github.com/eford/ApproximateBayesianComputing.jl} and \url{https://github.com/ExoJulia/ExoplanetsSysSim.jl} for ABC and SysSim respectively.} under the MIT Expat License.
We provide an overview of the the ABC methodology below and identify updates for this study.

\subsection{Approximate Bayesian Computing}
\label{sec:ABC}
In ABC, one uses a forward model to generate many synthetic data sets and a distance function to evaluate how similar the synthetic and observed data sets are.  
A simplistic implementation of ABC would involve drawing parameters for the forward model from their priors, generating a synthetic data set for each set of parameters, computing summary statistics for each synthetic data set and evaluating the distance function ($\rho(s_{\mathrm{obs}},s^*)$) between the summary statistics for each synthetic data set ($s^*$) and the summary statistics for the actual observations ($s_{\mathrm{obs}}$).  
The empirical distribution of the model parameters that result in synthetic data sets with a distance less than a distance threshold, $\epsilon$, constitutes the ABC posterior.  
In toy problems, one can often identify a set of sufficient statistics that completely specify the properties of the simulated data set.  
For example, if the synthetic data were draws from a normal distribution, then the sample mean and sample standard deviation of a sample would be sufficient statistics.  
By choosing a distance function that goes to zero only when the sufficient statistics for the synthetic and observed data sets are identical, one can prove that the ABC posterior approaches the true posterior as $\epsilon$ goes to zero.  
For finite $\epsilon$, the ABC posterior is broader than the true posterior, so the ABC posterior is conservative.  
In practice, ABC is most useful for complex problems where the observed data is not drawn from an analytical distribution, so it is not possible to identify sufficient statistics that completely describe the distribution of observations.  
In these cases, expert knowledge must be used to identify summary statistics and a distance function that capture the scientifically important features of the model and data. 

In this study, we adopt summary statistics and distance functions that have already been developed, tested and validated by \citet{HFR+2018} and \citet{HFR+2019}. 
For summary statistics, we set $s_k$ to the ratio of number of planet candidates detected (within the $k$th bin of orbital periods and planet sizes) to the number of target stars searched.  
We adopt a distance function
\begin{equation}
    \rho (s_{\mathrm{obs},k},s^*_k) = \sum_k \frac{|s_{\mathrm{obs},k} - s^*_k|}{\sqrt{s_{\mathrm{obs},k} + s^*_k}},
\end{equation}
where $s_{\mathrm{obs}, k}$ are summary statistics for \emph{Kepler} observations and $s^*_k$ are summary statistics for the synthetic catalog.

In order to improve sampling efficiency relative to the simplistic ABC implementation described above, we follow \citet{HFR+2018} and \citet{HFR+2019} and 
implement a Population Monte Carlo (PMC) sampler which effectively converges on the ABC posterior for the planet occurrence rates.  
The ABC-PMC algorithm is based on sequential importance sampling, where each ``generation'' consists of a set of ``particles''.  Each particle represents one set of the model parameters and a weight.\footnote{While one could make an analogy between particles and ``walkers'' in ensemble-based MCMC samplers, there are important technical distinctions.  For example, when using MCMC to sample from a posterior, a Markov chain approaches a fixed target distribution.  In ABC-PMC, the particles within each generation are drawn from a different sampling distribution.  The sampling distribution is adapted based on the previous generation, so particles from the final generation provide a useful approximation to the ABC posterior distribution.}
For the initial generation, particles are drawn from the prior and given equal weight.
For each subsequent generation, the last set of particles are used to construct a sampling density for proposing new particles.  
Proposals are accepted or rejected based on whether the resulting distance exceeds a distance threshold that is gradually decreased with subsequent generations.  
The ABC-PMC algorithm calculates weights for each accepted particle based on the ratio of the prior probability to the proposal density for that particle's parameter values, so as to ensure that a  kernel density estimator (KDE) constructed from the particles and their weights properly reflects the prior probability distribution.
The ABC-PMC algorithm terminates once the distance threshold is sufficiently small that additional iterations would not result in significant changes to the distribution of particles, using the criteria detailed in \citet{HFR+2018}.
The final set of particles is used to construct a KDE for the ABC posterior distribution for the model parameters.  
Notably, the ABC-PMC algorithm allows us to characterize the uncertainties in model parameters, including non-Gaussianity and asymmetries that are common when working with small sample sizes.  

For a more complete description of the ABC-PMC algorithm, as well as the results of extensive validation for characterizing exoplanet occurrence rates using our summary statistics and distance functions, see \cite{HFR+2018}. 
For completeness, we note the choice of the algorithmic parameters below.  
When applying ABC-PMC to the M dwarf catalog, we set the number of ``particles'' per generation to 500, larger than in \citet{HFR+2018} so as to ensure the tails of the ABC posteriors for each parameter are well characterized.  
As shown in \citet{HFR+2018}, it is important to generate simulated catalogs with the same number of target stars as in the selected sample of target stars, in order to accurately estimate uncertainties.
Using a larger stellar sample for simulated catalogs could provide a more precise estimate of the expected value of detected planets per star, but would result in underestimating the expected sample variance.  
We set $\tau = 2$, where $\tau$ describes how much the proposal distribution is broadened relative to the sample variance of particles in the current generation) for all generations.  

In this study, the model parameters ($f_{i,j}$'s) are the planet occurrence rates for each bin in a 2-D grid over orbital period and planet radius.  The bin edges for this study are: $P = $ \{0.5, 1, 2, 4, 8, 16, 32, 64, 128, 256\} days \& $R_p = $ \{0.5, 1, 1.5, 2, 2.5, 3, 4\} R$_{\oplus}$.  Note that in comparison to the grid used in \citet{HFR+2019} the longest period bin ($P = $ \{256, 500\}) and the largest radii bins ($R_p = $ \{4, 6, 8, 12, 16\}) have been removed. This choice was made due to the paucity of planet candidates within those ranges.  Radius bins of $0.25~\textrm{R}_\oplus$ were merged into bins of $0.5~\textrm{R}_\oplus$ due to the smaller number of targets and planet candidates in the M dwarf sample.

Next, we detail aspects of the forward model we considered and changed to account for the different stellar type being analyzed.

\subsection{Stellar Properties}
\label{secImproveStars}
In \citet{HFR+2019}, we make use of the second \emph{Gaia} data release (DR2) \citep{GBV+2018} to provide improved stellar radii for our selection of FGK target stars.  In the case of our M dwarf sample, a significant fraction of the target stars do not have stellar radii provided by the \emph{Gaia} catalog.

In order to get accurate stellar parameters for our entire target catalog, we make use of empirical relations from \citet{MFG+2015} between the 2MASS $Ks$ absolute magnitude ($M_{Ks}$) and stellar radius/mass. 
These relations were fit using 183 K7-M7 single stars\footnote{Since these empirical relations were fit to single stars, they do not account for stellar binarity. We assume that our selected M dwarf targets are single stars.}, with the relations showing scatter of 2-3\%. 
We cross-matched the \emph{Kepler} target list with \emph{Gaia} DR2 and use the \emph{Gaia} to 2MASS PSC cross-match results of \citet{MMF+2019}.  
We estimate $M_{Ks}$ by taking the (apparent) $Ks$ from 2MASS PSC and applying the distance modulus using the \emph{Gaia} DR2 measured parallax.  
Uncertainties from 2MASS $Ks$ and \emph{Gaia} DR2 parallaxes were propagated to estimate uncertainties on $M_{Ks}$. 
We also make use of 2MASS $J$ and $Ks$ as a color threshold to identify our target sample of M dwarfs (and some late K dwarfs).  
We discuss our use of the 2MASS PSC to select our M dwarf sample in more detail in \S\ref{secCat}.

Stellar radii estimates from \citet{MFG+2015} are typically larger than radii estimates listed in the \emph{Kepler} DR25 stellar catalog. 
This discrepancy between model and empirically derived stellar radii has been previously attributed to spots and magnetic activity of M dwarfs \citep[e.g.,][]{MM2001}. 
The underinflation of model M dwarf radii is also noted more recently in the \citet{MDM+2017} study that investigated eight nearby M dwarf systems with confirmed exoplanets discovered by \emph{Kepler}.  That study concludes that because Kepler-42 (an inactive, metal-poor star) was also inflated relative to model predictions, activity alone may not explain the discrepancy.

\subsection{Planet Detection Efficiency}  
\label{secDetEff}
Similar to \citet{HFR+2019}, we use a combined model for the probability of a transiting planet being detected and labeled as a planet candidate by the robovetter using the pixel-level transit injection tests \citep{C2017Pixel} and the associated robovetter results \citep{C2017Robo}:  
\begin{equation}
p_{\mathrm{det\&vet}}(SNR,N_{tr}) = c_{N_{tr}} \times \gamma(\alpha_{N_{tr}},\beta_{N_{tr}} \times SNR)/\Gamma(\alpha_{N_{tr}}),
\label{eqnDetAndVet}
\end{equation}
where $N_{tr}$ is the number of ``valid'' transits observed by \emph{Kepler}, $SNR$ is the estimated signal-to-noise ratio for the transit detection,  $\Gamma$ is the Gamma function defined as $\Gamma(z) = \int_{0}^{\infty}x^{z-1}e^{-x}\,dx$, $\gamma$ is the lower incomplete gamma function defined as $\gamma(s,x) = \int_{0}^{x}t^{s-1}e^{-t}\,dt$, and values for $\alpha$, $\beta$, and $c$ are given in Table \ref{tab:detvetmodel}.
Eqn. \ref{eqnDetAndVet} can be interpreted as the cumulative distribution function (CDF) for a Gamma distribution, scaled by $c_{N_{tr}}$.  This Gamma CDF parameterization of the detection efficiency curve was originally chosen due to its shape being a good fit to injection test results and published by \citet{CCB+2015}.  Subsequent studies \citep[e.g.,][]{HFR+2018,HFR+2019} have found that this parameterization continues to be useful when constructing a more detailed detection efficiency model that accounts for the number of transits observed.
\begin{table}
\caption[Parameters for $p_{\mathrm{det \& vet}}$]{Parameters for $p_{\mathrm{det \& vet}}$\label{tab:detvetmodel}}
\begin{tabular}{rrrr}
\multicolumn{1}{c}{$N_{tr}$}&\multicolumn{1}{c}{$\alpha$}&\multicolumn{1}{c}{$\beta$}&\multicolumn{1}{c}{$c$}\\
\hline\hline
 3       & 33.3884& 0.264472& 0.699093\\
 4       & 32.8860& 0.269577& 0.768366\\
 5       & 31.5196& 0.282741& 0.833673\\
 6       & 30.9919& 0.286979& 0.859865\\
 7-9     & 30.1906& 0.294688& 0.875042\\
 10-18   & 31.6342& 0.279425& 0.886144\\
 19-36   & 32.6448& 0.268898& 0.889724\\
 $\ge37$ & 27.8185& 0.32432 & 0.945075\\
\end{tabular}
\end{table}

\begin{figure}
\includegraphics[scale=0.52]{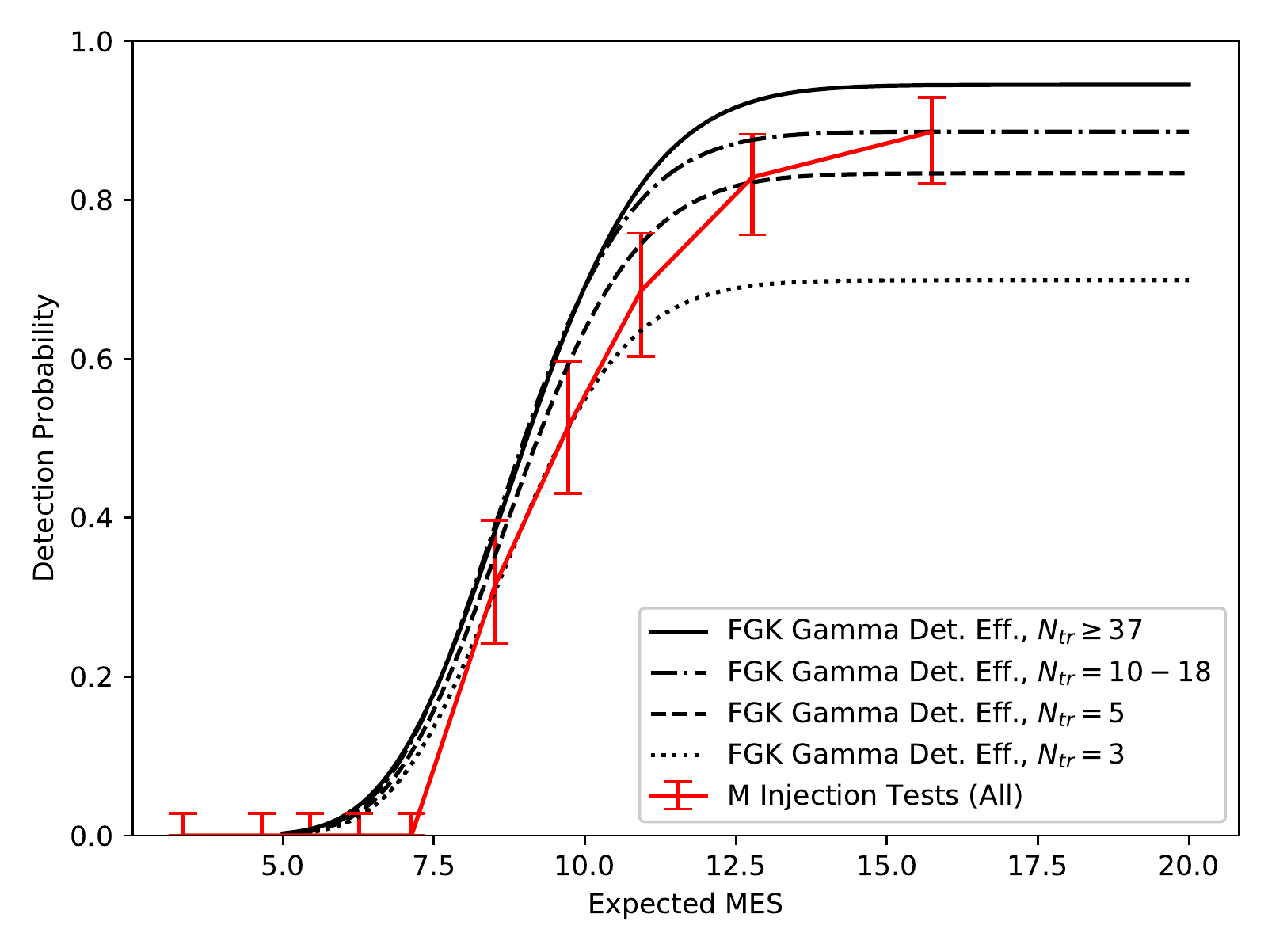}
\caption{Transit injection results for the subset of our M dwarf sample tested by \citet{C2017Pixel} (red) compared to four different detection efficiency curves (black) used in \citet{HFR+2019} for FGK dwarfs covering different ranges of expected number of transits ($N_{tr}$).  The M dwarf injection test results are reasonably bracketed by the FGK detection efficiency curves.
}
\label{figMInjTest}
\end{figure}

A more accurate characterization of planet detection for M dwarfs would require performing many more planet injection tests similar to \citet{C2017Pixel}, but focusing on M dwarf targets.
In particular, one would need to: (1) generate tens of thousands of synthetic data sets where simulated planetary signals are injected into actual \emph{Kepler} observations of M dwarf targets, so as to have realistic stellar noise properties; (2) analyze the simulated pixel-level light curves with the actual \emph{Kepler} pipeline; (3) apply the robovetter to the resulting data products to determine which simulated planets would have been detected as planet candidates; and (4) fit a new detection and vetting efficiency model to the results for which planets are detected as a function of $SNR$ and $N_{tr}$.
Unfortunately, the number of transit injections into M dwarf targets by \citet{C2017Pixel} is too small to accurately fit a new model for most ranges of $N_{tr}$.  However, comparing the injection test results for our M dwarf sample to the FGK detection model curve shows the two are consistent with one another as shown in Fig. \ref{figMInjTest}. 
Given this result, we have chosen to apply the FGK detection model directly to our M dwarf catalog.

\subsection{Model Parameterization}
\label{secImproveParam}
For our simulations, we perform five independent runs for each period bin that simultaneously fit seven radius bins over the ranges: $R_p = $ \{0.25, 0.5, 1, 1.5, 2, 2.5, 3, 4\} R$_{\oplus}$.  We have previously verified that seven bins simultaneously fit can still produce accurate estimates to within $10\%$ of the true occurrence rate (see Fig. 1 of \citet{HFR+2019}).  We discard the smallest radius bin due to a known bias that overestimates the rate in edge bins when incorporating stellar parameter uncertainties (see \S 2.5 from \citep{HFR+2019}).  
We perform five ABC-PMC calculations and calculate the relevant percentile (e.g., 50\% for bins with $\geq 2$ planet candidates) for each of the five simulations. We then report the median value of the five estimates as the final estimated rate.
Upper (lower) uncertainties indicate the difference between the 15.87th (84.13th) percentiles and the 50th percentile of the posterior produced by the simulation that gives the reported rate.  
In bins with zero or one planet candidates, we report the 84.13\% percentile.  

In \citet{HFR+2019} we showed that some occurrence rates (particularly those for small, long-period planets) were sensitive to the choice of prior.  
Therefore, we analyze the M dwarf sample using two different priors.
The two priors use two different model parameterizations, both chosen to ease interpretation of results.
Neither prior assumes a correlation between occurrence rates in different bins {\em a priori}.
Yet, the hierarchical model still allows for the posterior distributions for the intrinsic occurrence rates to have correlations. 
Our forward model accounts for the expected correlation in the observed rates of exoplanets for neighboring bins comes from imprecise determinations of planet radius that can result in a planet candidate being assigned to the incorrect radius bin.  
A correlation in the observed occurrence rates for two bins that is stronger than expected based on the uncertainties in radius measurements would lead to correlations in the posterior distributions for the corresponding occurrence rates.

For the first prior and model parameterization, we assume that each period range has a total occurrence rate drawn from a uniform prior, and assign a fraction of those planets to each radius bin.
The total occurrence rate for planets within a given range of orbital periods is treated as one model parameter, and the rates occurrence rates for each radius bin are constrained so they sum to the total rate for the corresponding period bin. 
For each bin in orbital period (indexed by $j$), we infer: 1) $f_{\mathrm{tot},j}$, the total occurrence rate for all planets within the $j$th range of orbital periods and the full range of planet sizes being considered, and 2) a vector $f_{\mathrm{rel},i,j}$: the relative occurrence rates, or more precisely the fractions of planets in the $j$th orbital period bin that have sizes falling within the range of each of the $i$th radius bins.
Thus, the occurrence rate for planets in the $i$th radius bin and $j$th period bin is given by $f_{i,j} = f_{\mathrm{tot,j}} f_{\mathrm{rel},i,j}$.
The upper limit for the uniform prior over $f_{\mathrm{tot,j}}$ is given by 
$f_{\mathrm{\max,tot,j}}=3\times \log(P_{\max,j}/P_{\min,j})/\log(2)$.
This prior choice is motivated by the desire for a minimally informative prior on the planet occurrence rate marginalized over planet size and the assumption that long-term orbital stability limits the total number of planets within a given period range.

From a theoretical perspective, N-body simulations show that systems with three or more planets typically go unstable if their initial semi-major axes are not sufficiently well separated  (e.g., \citet{OVT2017} and references therein).  
Given the chaotic nature of planetary systems and the large number of parameters for a three planet system, quantitative stability criteria are only available for special cases.  Therefore, it is useful to look at the distribution of orbital period ratios among planetary systems identified by {\emph Kepler} with at least three planets.  
For each set of three apparently adjacent planets, we compute the ratio of the orbital period of the outermost planet to orbital period of the innermost planet.  We find that this ratio is less than two for only 5\% (13 out of 244) of systems.  For the unusually closely packed triples, the minimum ratio is 1.42 and the median ratio is 1.71.  Our choice of prior reflects such systems being rare.
Note that our prior does not strictly exclude rare systems with more than three planets with orbital periods within a factor two of each other.  
Instead, the prior excludes the rate of planets being so high that every star would have 3 planets within a single period bin (i.e., a factor of two of each other). 

For each $f_{\mathrm{rel},i,j}$ (a vector with $j$ fixed), we then adopt a Dirichlet prior with concentration parameters proportional to $\log(R_{\max,i}/R_{\min,i})$ with the parameters normalized so that the smallest parameter is unity.   
The Dirichlet (or multivariate beta) distribution is a conjugate prior for the multinomial distribution, facilitating the interpretability of the model.  
For fitting seven radius bins, setting each concentration parameter to unity would result in a assigning equal prior probability to each point on the 6-simplex (i.e., generalization of a triangle to 6 dimensions, so the sum of the seven $f_{\mathrm{rel}}$'s is constrained to be unity) for relative occurrence rates at a given period.
Using equal concentration parameters would result in the expected value for the number of planets in each radius bin being identical.  
In contrast, our choice of concentration parameters results in a prior where the expected value for the number of planets in each radius bin is proportional to $\log(R_{\max,i}/R_{\min,i})$.  
Setting the smallest concentration parameter to one prevents the prior from becoming peaked at any corner of the 6-simplex for the relative rates (i.e., avoids favoring all the planets being assigned to a single radius bin).

For the second prior and model parameterization, we assign each bin in period and radius its own occurrence rate, $f_{i,j}$.  
This prior choice makes no assumptions about smoothness of the planet occurrence rate as a function of size or distance.  It was selected to facilitate ease of interpretation and to estimate occurrence rates using a minimal number of prior assumptions possible given our 2-D period-radius binned occurrence rate parameterization. 
We adopt a uniform prior for $f_{i,j}$ over $[0,f_{\mathrm{\max,i,j}})$.
The upper limit was set to 
$f_{\mathrm{\max,i,j}}=C\times \log(P_{\max,j}/P_{\min,j})/\log(2)\times \log(R_{\max,i}/R_{\min,i})/\log(2)$, 
with $C = 2$ in most cases.\footnote{When we perform calculations over a period-radius grid with equivalent flux to the \citet{HFR+2019} grid for FGK targets, we instead use $C = 1.5$.  See \S\ref{secCompFGK} for details on how this equivalent flux period-radius grid was determined.}

While a uniform prior for each radius bin may seem to be making minimal assumptions, the implied prior for the sum of the occurrence rates within a given period range is not uniform, but rather peaked near $\sum_j f_{\mathrm{\max,i,j}}/2$.  
This can be seen more simply by considering the sum of seven random variables (i.e., our number of radius bins for each period range), each drawn independently from a uniform distribution between zero and one.  The mean of the resulting distribution is $\frac{7}{2}$ and the standard deviation is $\sqrt{\frac{7}{12}}$, just 22\% of the mean.  In contrast, a uniform distribution between zero and seven has the same mean, but a standard deviation 2.6 times greater.  
Thus, choosing a uniform prior for each bin runs the risk of the prior exerting an undesirable influence on the posterior distribution for occurrence rates marginalized over planet size, particularly when working with a relatively small sample of {bf M dwarfs}.

Given the above considerations, we consider results using the Dirichlet prior to be the baseline results for this study.  
Nevertheless, the results based on uniform priors for each occurrence rate, provide a useful contrast that can help assess the sensitivity of our result to the choice of prior choice.

Studies investigating the architecture of inner planetary systems around sun-like stars suggest that the typical inner planetary system contains multiple planets \citep[e.g.,][]{MPA+2018,ZPW+2018,HeFR2019,SKC2019,ZCH2019}.  Studies focusing on the architectures of M dwarf planetary systems are also suggestive of multiple planets being common \citep[e.g.,][]{MMV+2015,BJ2016}, but have made additional assumptions in order to perform inference with the much smaller sample of planetary systems observed around M dwarfs.  
In this study, we focus on the question of inferring average number of planets (within a given range of sizes and period) per star and not the average number of planets per planetary system or the fraction of stars with planets.  Fortunately, inferences about the average number of planets per star are not affected by whether a given set of planets are orbiting a single star or spread over a set of stars (with the same stellar properties).  Thus, the distribution for the number of planets per planetary system and the distribution of mutual inclinations of planets in a given planetary system do not affect our median estimated rates. The existence of correlations, however, could lead to underestimates of the uncertainties of our rates.

Previous studies of planets around sun-like stars have also found that the sizes of planets within a planetary system are correlated with each other \citep[e.g.,][]{MWL2017,HeFR2019,WP2020}.  Such a correlation may also be present for planets around M stars.  If present, such correlations would not affect the inferences for the occurrence rate per star in our study, but it could also lead to the uncertainty in occurrence rates  being underestimated. 
Investigating architectural questions such as those described above would require a model that includes parameters to characterize the architectural properties of planetary systems and either a likelihood or distance function that makes use of information about the observed properties of multiple planet systems \citep[e.g.,][]{MPA+2018,HeFR2019}.  Such an analysis is beyond the scope of this study.

\section{Application to DR25 M dwarfs}
\label{secResults}
\begin{figure}
\includegraphics[scale=0.56]{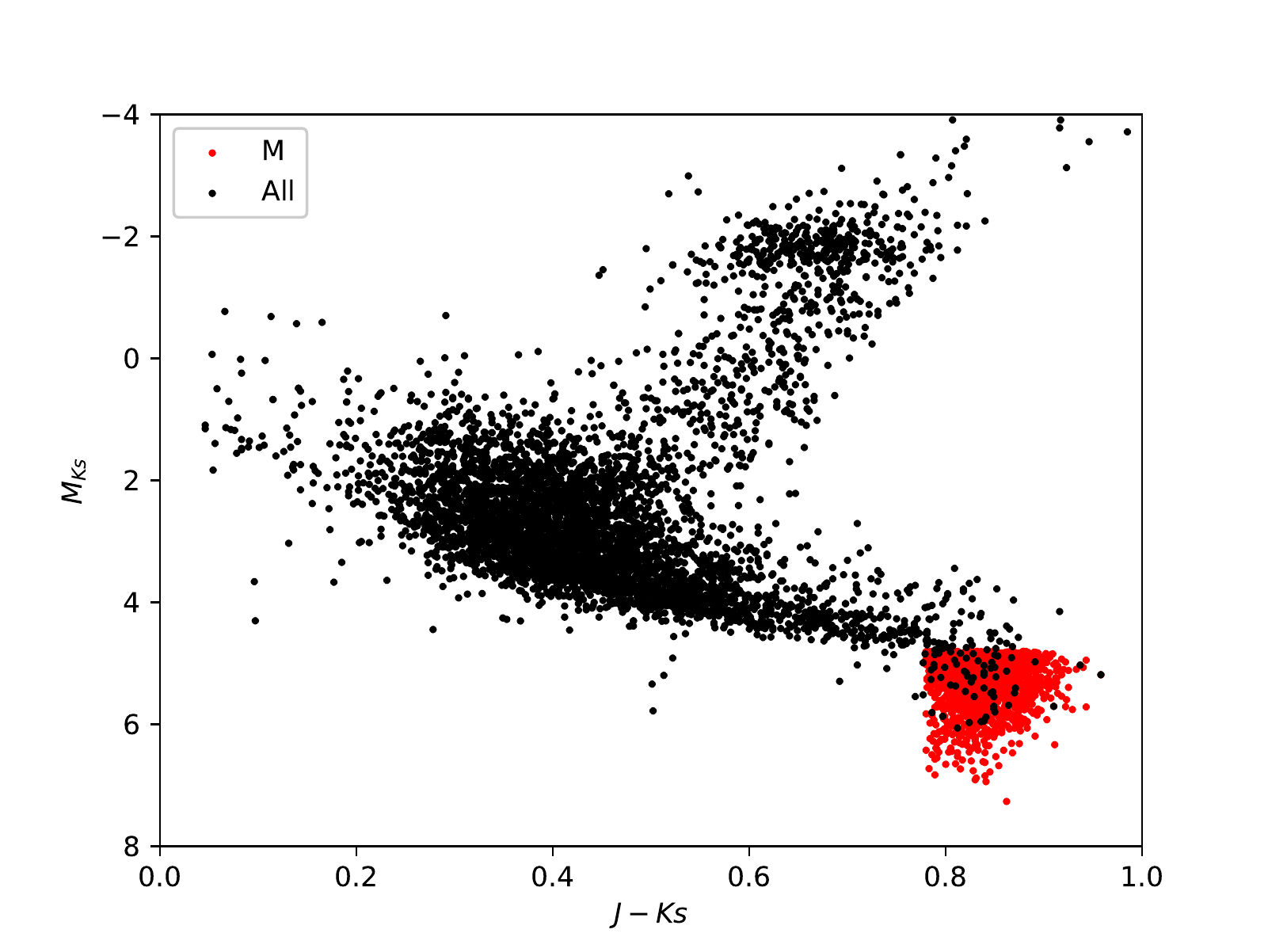}
\includegraphics[scale=0.56]{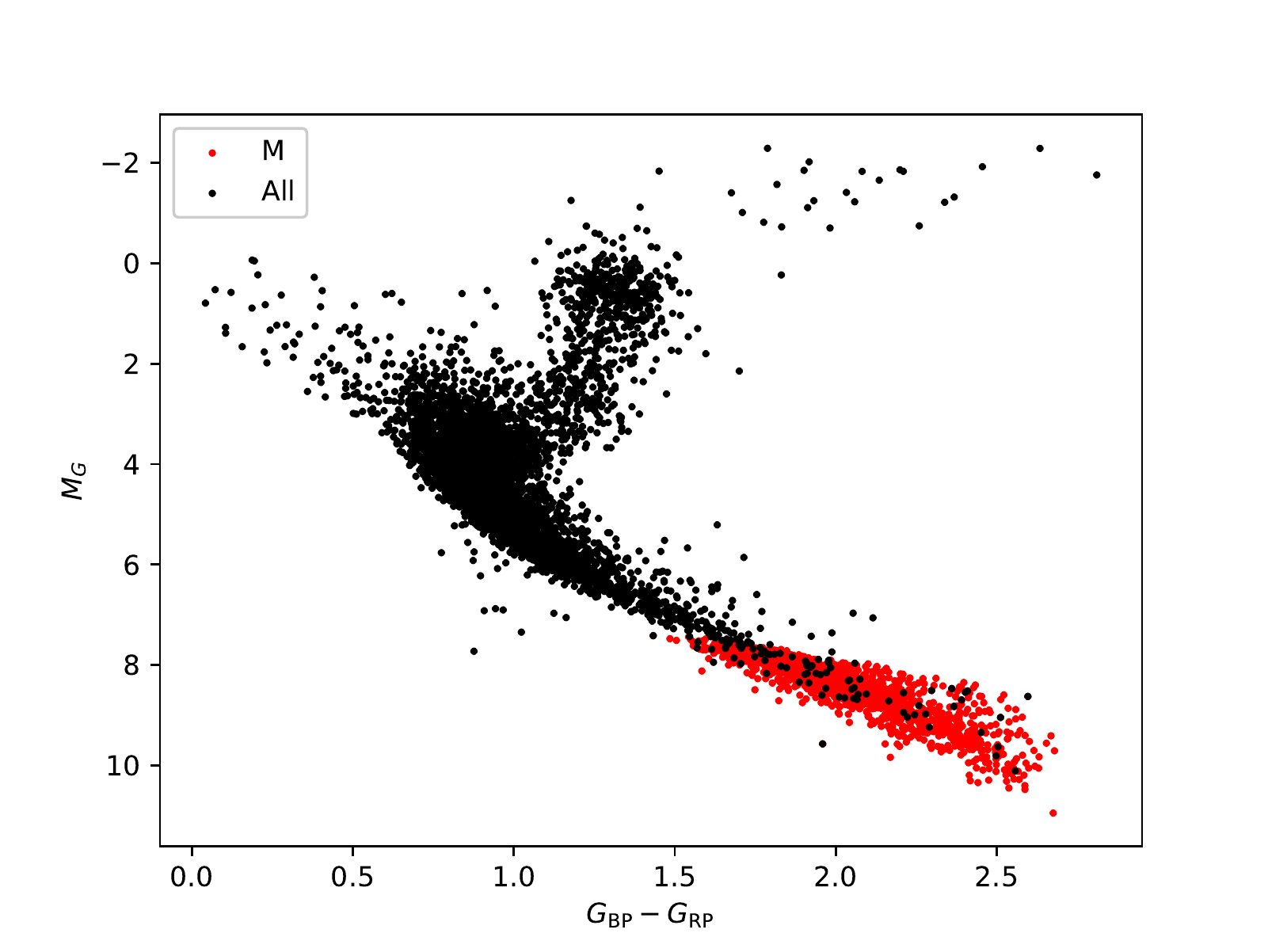}
\caption{Our full M dwarf target sample (red points) compared to 5000 randomly sampled \emph{Kepler} DR25-\emph{Gaia} DR2 cross-matched targets that pass our photometry quality cuts (Criteria 1-5; black points).  The axes in the top panel depict $M_{Ks}$ and $J-Ks$ taken from 2MASS measurements.  The axes in the bottom panel depict $M_G$ and $G_\textrm{BP} - G_\textrm{RP}$ which are derived from \emph{Gaia} measurements. 
$M_G$ does not include a correction using the inferred extinction from \emph{Gaia}'s Apsis module \citep{AFC+2018}, as these M dwarfs are quite nearby and have minimal extinction.
}
\label{figMSamp}
\end{figure}

\subsection{Catalog Selection}
\label{secCat}
In \citet{HFR+2019}, we made use of the second \emph{Gaia} data release (DR2) \citep{GBV+2018} to acquire significantly improved stellar parameters for the overwhelming majority of the \emph{Kepler} target stars.  In the case of M dwarfs, however, we lose a significant fraction of stars if we require the \emph{Gaia} DR2 to have an estimated stellar radius for every target.  To that end, we incorporate $J$ and $Ks$ magnitude measurements from the 2MASS Point Source Catalog \citep{SCS+2006} adopting the best neighbor cross match between \emph{Gaia} DR2 and 2MASS performed by the \emph{Gaia} mission team \citep{MMF+2019}.
We use \emph{Gaia}'s precise parallax measurements and the 2MASS $J$ and $Ks$ magnitudes to accurately identify main-sequence M dwarfs based on their position in the color-luminosity diagram. 

Our stellar catalog stars begins with a foundation of \emph{Kepler} target stars.  Then, the \emph{Kepler} DR25 stellar properties catalog is augmented with data from \emph{Gaia} DR2 \citep{GBV+2018} and 2MASS PSC \citep{SCS+2006}.
We use many similar selection criteria to select main-sequence M stars in this study as we did in \citet{HFR+2019}.  
For completeness, we list the full selection criteria below (in order of application), along  with explanations for additions and changes from our previous study:

\begin{enumerate}
\item \textbf{Require that the \emph{Kepler} magnitude and the \emph{Gaia} $G$ magnitude are consistent.}  The \emph{Kepler} and \emph{Gaia} $G$ bandpasses have a very high degree of overlap, so correctly cross-matched targets should have $K_\textrm{p}$ and $G$ magnitudes that are very similar.  Indeed, the distribution of $K_\textrm{p}-G$ for cross-matched target stars appears nearly Gaussian and centered close to zero.  We reject targets if their $K_\textrm{p}-G$  magnitude deviates by more than 1.5$\sigma$ of the median of the $K_\textrm{p}-G$  distribution for the cross matched \emph{Kepler} and \emph{Gaia} catalogs where $\sigma$ is the standard deviation of the $K_\textrm{p}-G$ distribution.
\item \textbf{Require \emph{Gaia} Priam processing flags that indicate the parallax value is strictly positive and both colors are close to the standard locus for main-sequence stars}.  
This rejects \emph{Kepler} targets which are unlikely to be main sequence stars or are so distant that \emph{Gaia} does not have a good parallax measurement and the stellar radius will be highly uncertain.  
\item \textbf{Require \emph{Gaia} parallax error is less than $30\%$ the parallax value}.
We relax our error tolerance for the \emph{Gaia} parallax relative to \citet{HFR+2019}, so as to increase the completeness of our M dwarf sample Most M dwarf targets are faint in the \emph{Gaia} bands, so their parallax uncertainties are expected to be larger than for similar distance FGK stars.
\item \textbf{Require that \emph{Kepler} DR 25 provide \emph{Kepler} data span, duty cycle, and limb-darkening coefficients}.
\item \textbf{Require that the \emph{Kepler} target was observed for $>4$ quarters and must have been on the Exoplanet target list for at least one quarter}. The majority of target stars were observed by \emph{Kepler} for exoplanet identification, but some stars were selected for observation for secondary goals (e.g. asteroseismology). We choose to exclude those stars not explicitly part of the exoplanet search target list.
\item \textbf{Require the target to have a color $J-Ks \geq 0.779$ from 2MASS photometry}.  
This color cut results in selecting cool stars (M stars and red giants) and is more accurate than using the temperature from the \emph{Kepler} Input Catalog or the \emph{Gaia} color $G_\textrm{BP} - G_\textrm{RP}$.  The color threshold corresponds to the K7 stellar type color for the main sequence fit described by \citet{M2019, PM2013}.
\item \textbf{Require the target star to have an absolute magnitude, $M_{Ks} \geq 4.8$}.
This $M_{Ks}$ threshold corresponds to the K7 stellar type for the main sequence fit described by \citet{M2019, PM2013}. We compute $M_{Ks}$ by taking the 2MASS measured apparent magnitude $Ks$ and applying the distance modulus using the \emph{Gaia} DR2 parallax.  Given that all M dwarfs in the \emph{Kepler} target list must be nearby, we expect dust to have a negligible effect on the magnitude measurement and neglect extinction.  \citet{RCA+2011} previously found a color excess of $E_{b-y} < 0.07$ in the direction of the \emph{Kepler} field within 500 pc. Only one \emph{Kepler} target in our final M dwarf catalog has a distance estimated using the \emph{Gaia} DR2 parallax that is larger than 500 pc, KIC 10020298 at $\approx514$ pc. Adopting reddening coefficients $R$ of 4.3 for $R(b-y)$ (used by \citet{GMK+2016} in their extinction analysis in Appendix C), we expect $E_{B-V} \ll 0.016$.  Taking $R(Ks) = 0.306$ from \citet{YLX2013}, we then arrive at an extinction $A(Ks) \ll 0.005$ that is significantly less than the minimum uncertainty of $0.11$ on $Ks$ for the selected targets.
\end{enumerate}

Using these cuts we arrive at a final M dwarf sample of 1,746 targets with 89 associated planet candidates within our period-radius grid.  In Fig. \ref{figMSamp} we compare our M dwarf population to the full \emph{Kepler} population.
We find that our sample selection includes $\sim70\%$ of the Kepler targets with $G_\textrm{BP}-G_\textrm{RP} > 1.5$ and $M_G>7.5$. 
In comparison with the \citet{DC2015} catalog of 2,543 presumed M dwarf targets, we find a overlap of approximately 1,000 targets.  We note that \citet{DC2015} initially uses $T_{eff}$ and $\log g$ measurements from the obsolete Q1-Q16 \emph{Kepler} stellar catalog, which are less accurate than 2MASS measurements in combination with \emph{Gaia} parallaxes.

\subsection{Planet Occurrence Rates}
\label{secRates}
\begin{figure*}
\includegraphics[scale=0.43]{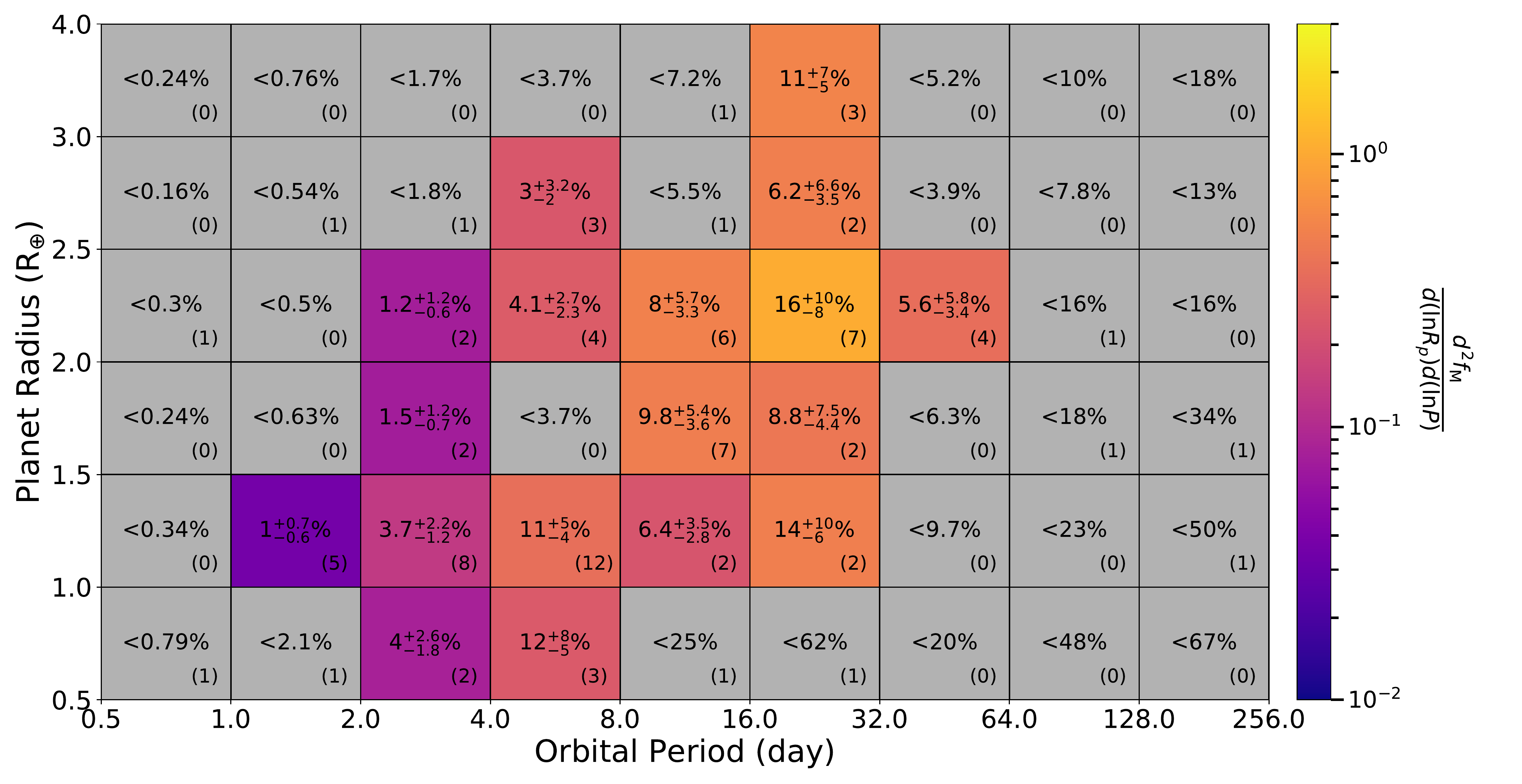}
\includegraphics[scale=0.43]{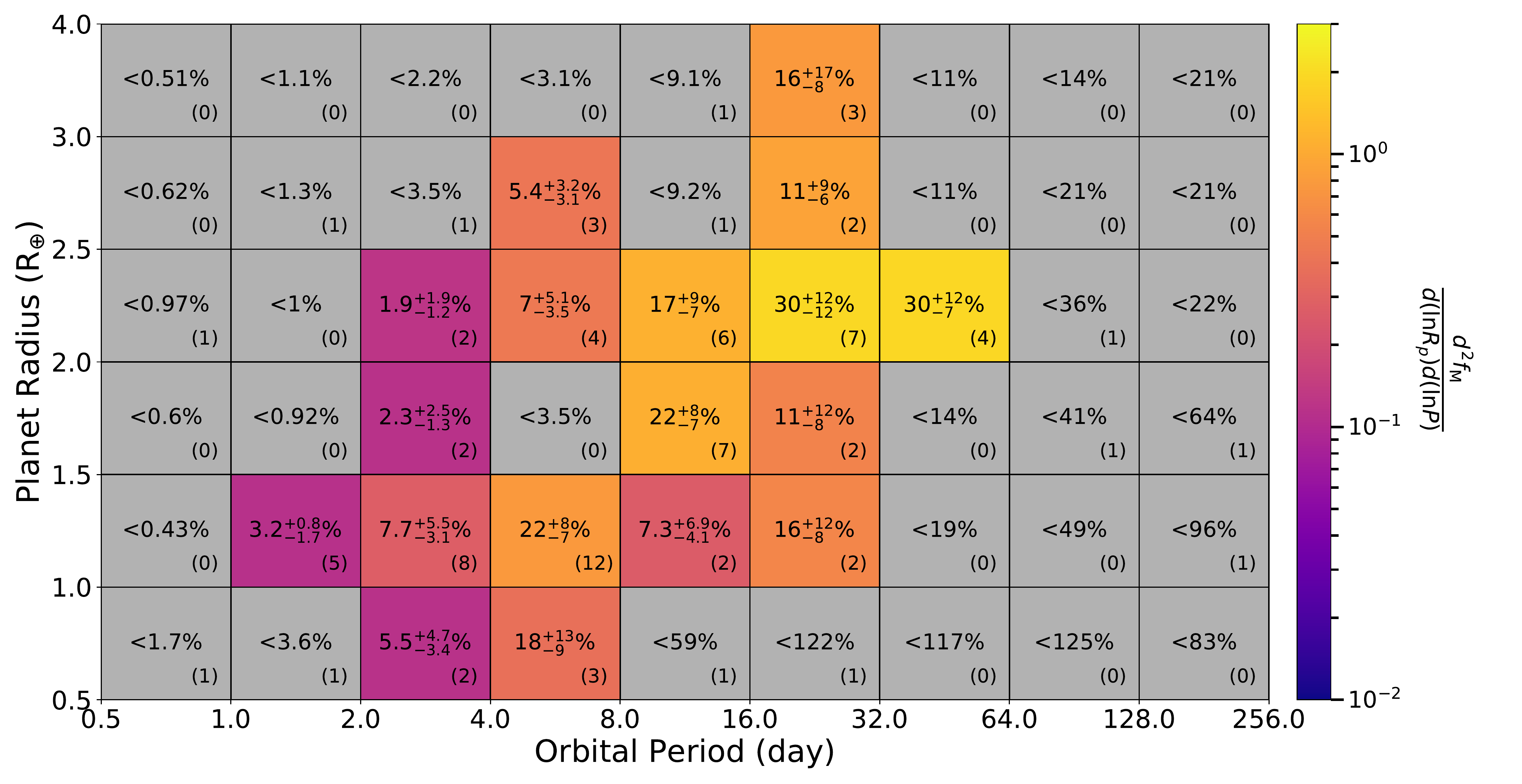}
\caption{Inferred occurrence rates for \emph{Kepler}'s DR25 planet candidates associated with high-quality M target stars with two choices of prior: uniform prior on total rate over each period range and Dirichlet prior on fraction of total rate assigned to each radius bin (top panel); and independent uniform priors per bin (bottom panel).  
These rates are based on a combined detection and vetting efficiency model that was fit to flux-level planet injection tests.  
The numerical values of the occurrence rates ($f_\textrm{M}$) are stated as percentage (i.e., $10^{-2}$). 
The uncertainties shown are the differences between the median and the 15.87th or 84.13th percentile.
The color coding of each cell is based on $(d^{2}f_\mathrm{M})/[d (\ln{R_{p}})~d(\ln{P})]$, which provides an occurrence rate normalized to the width of the bin and therefore is not dependent on choice of grid density.
Cells colored gray have estimated upper limits for the occurrence rate.
The number of DR25 planet candidates within each bin is listed in parentheses in the bottom right corner of each bin.
Note that the bin sizes are not constant.
The upper limits for planets of 3-4$R_\oplus$ also include the upper limits for larger planets (for the same period range and stellar sample), since the \emph{Kepler} sample contains no larger planets transiting M dwarfs and the detection efficiency for planets of this size is already high.  Any modest increase in the detection efficiency would only result in more strict upper limits on the occurrence rate of larger planets.}
\label{figRates}
\end{figure*}

In Table \ref{tab:occ_rates} we list the derived occurrence rates using our M dwarf sample for two different choices of prior as described in \S\ref{secImproveParam}.  We also show the derived occurrence rates in Fig. \ref{figRates}.

A bin's rate estimated using the uniform prior parameterization is generally larger than the same bin's rate estimated using the Dirichlet prior parameterization by a factor of $\simeq$2.
We caution that the actual ratio ranges from 1.1 to 5.2 depending on the specific bin, so we recommend referring to the specific result for each bin.
We attribute this difference to the different approaches the two priors take to the total occurrence rate over multiple radius bins with the same period range.
The Dirichlet prior parameterization assumes a uniform prior over the total rate. In contrast, the uniform prior assumes that the true rate of planets is independent between all bins, which means that the total rate follows an Irwin-Hall distribution. The prior distribution of each individual bin is therefore peaked at a larger occurrence rate when using the uniform prior than when using the Dirichlet prior, generally by a factor of $\sim 3-5$.
As expected, the ratio of the estimated rates with the two prior choices is less extreme, since it combines information from the prior distributions and the constraints of the observed Kepler planet candidate data.
The Dirichlet prior also has a broad tail towards rates smaller than its peak value in contrast to the symmetric uniform prior.
The combination of these differences in prior distribution and the constraints of the observed Kepler planet candidate data produce the smaller average factor of 2 difference in estimated rates resulting from the two prior choices.

\begin{table*}
\caption{DR25 M Dwarf Planet Occurrence Rates and Ratios with FGK Rates}
\label{tab:occ_rates}
\begin{tabular}{rrrrrrr}
\hline
\multicolumn{1}{c}{Period}&
\multicolumn{1}{c}{Radius}&
\multicolumn{1}{c}{Number of DR25}&
\multicolumn{2}{c}{Dirichlet Prior}&
\multicolumn{2}{c}{Uniform Prior}\\
\multicolumn{1}{c}{(days)}&
\multicolumn{1}{c}{(R$_{\oplus}$)}&
\multicolumn{1}{c}{Planet Candidates}&
\multicolumn{1}{c}{$f_\textrm{M}$}&
\multicolumn{1}{c}{$f_\textrm{M}/f_\textrm{FGK}$}&
\multicolumn{1}{c}{$f_\textrm{M}$}&
\multicolumn{1}{c}{$f_\textrm{M}/f_\textrm{FGK}$}\\
\hline
$\phn\phn 0.50-\phn\phn 1.00$&$\phn0.50-\phn1.00$&$1$&$<7.9\times10^{-3}$&$<2.5$&$<1.7\times10^{-2}$&$<6.4$\\
$\phn\phn 0.50-\phn\phn 1.00$&$\phn1.00-\phn1.50$&$0$&$<3.4\times10^{-3}$&$<2.7$&$<4.3\times10^{-3}$&$<3.4$\\
$\phn\phn 0.50-\phn\phn 1.00$&$\phn1.50-\phn2.00$&$0$&$<2.4\times10^{-3}$&$<2.2$&$<6.0\times10^{-3}$&$<6.7$\\
$\phn\phn 0.50-\phn\phn 1.00$&$\phn2.00-\phn2.50$&$1$&$<3.0\times10^{-3}$&N/A&$<9.7\times10^{-3}$&N/A\\
$\phn\phn 0.50-\phn\phn 1.00$&$\phn2.50-\phn3.00$&$0$&$<1.6\times10^{-3}$&N/A&$<6.2\times10^{-3}$&N/A\\
$\phn\phn 0.50-\phn\phn 1.00$&$\phn3.00-\phn4.00$&$0$&$<2.4\times10^{-3}$&N/A&$<5.1\times10^{-3}$&N/A\\
\hline
$\phn\phn 1.00-\phn\phn 2.00$&$\phn0.50-\phn1.00$&$1$&$<2.1\times10^{-2}$&$<4.3$&$<3.6\times10^{-2}$&$<7.8$\\
$\phn\phn 1.00-\phn\phn 2.00$&$\phn1.00-\phn1.50$&$5$&$1.02^{+0.70}_{-0.57}\times10^{-2}$&$3.0^{+2.8}_{-1.7}$&$3.18^{+0.77}_{-1.66}\times10^{-2}$&$8.6^{+10.3}_{-4.2}$\\
$\phn\phn 1.00-\phn\phn 2.00$&$\phn1.50-\phn2.00$&$0$&$<6.3\times10^{-3}$&$<4.7$&$<9.2\times10^{-3}$&$<8.0$\\
$\phn\phn 1.00-\phn\phn 2.00$&$\phn2.00-\phn2.50$&$0$&$<5.0\times10^{-3}$&$<34.7$&$<1.0\times10^{-2}$&$<93.2$\\
$\phn\phn 1.00-\phn\phn 2.00$&$\phn2.50-\phn3.00$&$1$&$<5.4\times10^{-3}$&$<38.8$&$<1.3\times10^{-2}$&$<107.4$\\
$\phn\phn 1.00-\phn\phn 2.00$&$\phn3.00-\phn4.00$&$0$&$<7.6\times10^{-3}$&N/A&$<1.1\times10^{-2}$&N/A\\
\hline
$\phn\phn 2.00-\phn\phn 4.00$&$\phn0.50-\phn1.00$&$2$&$4.0^{+2.6}_{-1.8}\times10^{-2}$&$1.1^{+0.8}_{-0.5}$&$5.5^{+4.7}_{-3.4}\times10^{-2}$&$1.4^{+1.5}_{-0.9}$\\
$\phn\phn 2.00-\phn\phn 4.00$&$\phn1.00-\phn1.50$&$8$&$3.7^{+2.2}_{-1.2}\times10^{-2}$&$3.2^{+2.2}_{-1.4}$&$7.7^{+5.5}_{-3.1}\times10^{-2}$&$6.4^{+4.5}_{-2.8}$\\
$\phn\phn 2.00-\phn\phn 4.00$&$\phn1.50-\phn2.00$&$2$&$1.50^{+1.20}_{-0.70}\times10^{-2}$&$1.7^{+1.8}_{-0.9}$&$2.3^{+2.5}_{-1.3}\times10^{-2}$&$2.8^{+2.7}_{-1.6}$\\
$\phn\phn 2.00-\phn\phn 4.00$&$\phn2.00-\phn2.50$&$2$&$1.20^{+1.23}_{-0.55}\times10^{-2}$&$5.1^{+7.3}_{-2.8}$&$1.9^{+1.9}_{-1.2}\times10^{-2}$&$10.1^{+17.1}_{-6.1}$\\
$\phn\phn 2.00-\phn\phn 4.00$&$\phn2.50-\phn3.00$&$1$&$<1.8\times10^{-2}$&$<37.5$&$<3.5\times10^{-2}$&$<82.6$\\
$\phn\phn 2.00-\phn\phn 4.00$&$\phn3.00-\phn4.00$&$0$&$<1.7\times10^{-2}$&$<21.0$&$<2.2\times10^{-2}$&$<29.6$\\
\hline
$\phn\phn 4.00-\phn\phn 8.00$&$\phn0.50-\phn1.00$&$3$&$1.19^{+0.84}_{-0.49}\times10^{-1}$&$0.9^{+0.8}_{-0.4}$&$1.84^{+1.31}_{-0.88}\times10^{-1}$&$1.6^{+1.3}_{-0.8}$\\
$\phn\phn 4.00-\phn\phn 8.00$&$\phn1.00-\phn1.50$&$12$&$1.05^{+0.49}_{-0.35}\times10^{-1}$&$2.5^{+1.4}_{-1.2}$&$2.20^{+0.75}_{-0.65}\times10^{-1}$&$5.8^{+3.6}_{-2.4}$\\
$\phn\phn 4.00-\phn\phn 8.00$&$\phn1.50-\phn2.00$&$0$&$<3.7\times10^{-2}$&$<1.7$&$<3.5\times10^{-2}$&$<1.7$\\
$\phn\phn 4.00-\phn\phn 8.00$&$\phn2.00-\phn2.50$&$4$&$4.1^{+2.7}_{-2.3}\times10^{-2}$&$3.6^{+3.4}_{-1.9}$&$7.0^{+5.1}_{-3.5}\times10^{-2}$&$7.8^{+6.0}_{-4.4}$\\
$\phn\phn 4.00-\phn\phn 8.00$&$\phn2.50-\phn3.00$&$3$&$3.0^{+3.2}_{-2.0}\times10^{-2}$&$2.5^{+3.1}_{-1.5}$&$5.4^{+3.2}_{-3.1}\times10^{-2}$&$5.5^{+4.7}_{-2.9}$\\
$\phn\phn 4.00-\phn\phn 8.00$&$\phn3.00-\phn4.00$&$0$&$<3.7\times10^{-2}$&$<17.1$&$<3.1\times10^{-2}$&$<19.4$\\
\hline
$\phn\phn 8.00-\phn16.00$&$\phn0.50-\phn1.00$&$1$&$<2.5\times10^{-1}$&$<2.3$&$<5.9\times10^{-1}$&$<6.0$\\
$\phn\phn 8.00-\phn16.00$&$\phn1.00-\phn1.50$&$2$&$6.4^{+3.5}_{-2.8}\times10^{-2}$&$1.0^{+0.8}_{-0.5}$&$7.3^{+6.9}_{-4.1}\times10^{-2}$&$1.0^{+1.2}_{-0.8}$\\
$\phn\phn 8.00-\phn16.00$&$\phn1.50-\phn2.00$&$7$&$9.8^{+5.4}_{-3.6}\times10^{-2}$&$3.4^{+2.8}_{-1.5}$&$2.18^{+0.82}_{-0.75}\times10^{-1}$&$7.2^{+4.3}_{-2.9}$\\
$\phn\phn 8.00-\phn16.00$&$\phn2.00-\phn2.50$&$6$&$8.0^{+5.7}_{-3.3}\times10^{-2}$&$2.7^{+2.2}_{-1.2}$&$1.73^{+0.95}_{-0.74}\times10^{-1}$&$5.3^{+4.0}_{-2.3}$\\
$\phn\phn 8.00-\phn16.00$&$\phn2.50-\phn3.00$&$1$&$<5.5\times10^{-2}$&$<2.2$&$<9.2\times10^{-2}$&$<3.4$\\
$\phn\phn 8.00-\phn16.00$&$\phn3.00-\phn4.00$&$1$&$<7.2\times10^{-2}$&$<12.0$&$<9.1\times10^{-2}$&$<17.2$\\
\hline
$\phn16.00-\phn32.00$&$\phn0.50-\phn1.00$&$1$&$<6.2\times10^{-1}$&$<4.1$&$<1.2\times10^{0}$&$<7.4$\\
$\phn16.00-\phn32.00$&$\phn1.00-\phn1.50$&$2$&$1.39^{+0.98}_{-0.60}\times10^{-1}$&$2.2^{+2.3}_{-1.1}$&$1.61^{+1.16}_{-0.84}\times10^{-1}$&$2.5^{+2.8}_{-1.5}$\\
$\phn16.00-\phn32.00$&$\phn1.50-\phn2.00$&$2$&$8.8^{+7.5}_{-4.4}\times10^{-2}$&$3.1^{+3.1}_{-1.8}$&$1.06^{+1.22}_{-0.75}\times10^{-1}$&$3.6^{+4.3}_{-2.2}$\\
$\phn16.00-\phn32.00$&$\phn2.00-\phn2.50$&$7$&$1.63^{+0.96}_{-0.80}\times10^{-1}$&$2.9^{+2.2}_{-1.3}$&$3.0^{+1.2}_{-1.2}\times10^{-1}$&$5.7^{+2.7}_{-2.4}$\\
$\phn16.00-\phn32.00$&$\phn2.50-\phn3.00$&$2$&$6.2^{+6.6}_{-3.5}\times10^{-2}$&$1.7^{+1.4}_{-1.1}$&$1.15^{+0.91}_{-0.62}\times10^{-1}$&$3.0^{+2.7}_{-2.0}$\\
$\phn16.00-\phn32.00$&$\phn3.00-\phn4.00$&$3$&$1.09^{+0.73}_{-0.47}\times10^{-1}$&$4.9^{+4.2}_{-2.7}$&$1.57^{+1.68}_{-0.82}\times10^{-1}$&$7.1^{+6.4}_{-3.9}$\\
\hline
$\phn32.00-\phn64.00$&$\phn0.50-\phn1.00$&$0$&$<2.0\times10^{-1}$&$<1.5$&$<1.2\times10^{0}$&$<8.7$\\
$\phn32.00-\phn64.00$&$\phn1.00-\phn1.50$&$0$&$<9.7\times10^{-2}$&$<1.5$&$<1.9\times10^{-1}$&$<3.6$\\
$\phn32.00-\phn64.00$&$\phn1.50-\phn2.00$&$0$&$<6.3\times10^{-2}$&$<2.1$&$<1.4\times10^{-1}$&$<4.6$\\
$\phn32.00-\phn64.00$&$\phn2.00-\phn2.50$&$4$&$5.6^{+5.8}_{-3.4}\times10^{-2}$&$1.2^{+1.7}_{-0.7}$&$2.95^{+1.18}_{-0.73}\times10^{-1}$&$6.3^{+4.6}_{-3.0}$\\
$\phn32.00-\phn64.00$&$\phn2.50-\phn3.00$&$0$&$<3.9\times10^{-2}$&$<0.6$&$<1.1\times10^{-1}$&$<1.9$\\
$\phn32.00-\phn64.00$&$\phn3.00-\phn4.00$&$0$&$<5.2\times10^{-2}$&$<2.8$&$<1.1\times10^{-1}$&$<5.8$\\
\hline
$\phn64.00-128.00$&$\phn0.50-\phn1.00$&$0$&$<4.8\times10^{-1}$&$<0.9$&$<1.3\times10^{0}$&$<2.3$\\
$\phn64.00-128.00$&$\phn1.00-\phn1.50$&$0$&$<2.3\times10^{-1}$&$<2.0$&$<4.9\times10^{-1}$&$<5.2$\\
$\phn64.00-128.00$&$\phn1.50-\phn2.00$&$1$&$<1.8\times10^{-1}$&$<4.5$&$<4.1\times10^{-1}$&$<10.2$\\
$\phn64.00-128.00$&$\phn2.00-\phn2.50$&$1$&$<1.6\times10^{-1}$&$<2.9$&$<3.6\times10^{-1}$&$<6.5$\\
$\phn64.00-128.00$&$\phn2.50-\phn3.00$&$0$&$<7.8\times10^{-2}$&$<2.7$&$<2.1\times10^{-1}$&$<6.8$\\
$\phn64.00-128.00$&$\phn3.00-\phn4.00$&$0$&$<10.0\times10^{-2}$&$<3.3$&$<1.4\times10^{-1}$&$<4.8$\\
\hline
$128.00-256.00$&$\phn0.50-\phn1.00$&$0$&$<6.7\times10^{-1}$&$<1.2$&$<8.3\times10^{-1}$&$<1.7$\\
$128.00-256.00$&$\phn1.00-\phn1.50$&$1$&$<5.0\times10^{-1}$&$<4.8$&$<9.6\times10^{-1}$&$<9.5$\\
$128.00-256.00$&$\phn1.50-\phn2.00$&$1$&$<3.4\times10^{-1}$&$<7.0$&$<6.4\times10^{-1}$&$<14.1$\\
$128.00-256.00$&$\phn2.00-\phn2.50$&$0$&$<1.6\times10^{-1}$&$<3.4$&$<2.2\times10^{-1}$&$<4.3$\\
$128.00-256.00$&$\phn2.50-\phn3.00$&$0$&$<1.3\times10^{-1}$&$<2.0$&$<2.1\times10^{-1}$&$<3.8$\\
$128.00-256.00$&$\phn3.00-\phn4.00$&$0$&$<1.8\times10^{-1}$&$<8.4$&$<2.1\times10^{-1}$&$<9.6$\\
\end{tabular}
\end{table*}

\section{Discussion}
\label{secDiscussion}

\subsection{Comparison to FGK Occurrence Rates}
\label{secCompFGK}
\begin{figure*}
\centering
\includegraphics[scale=0.43]{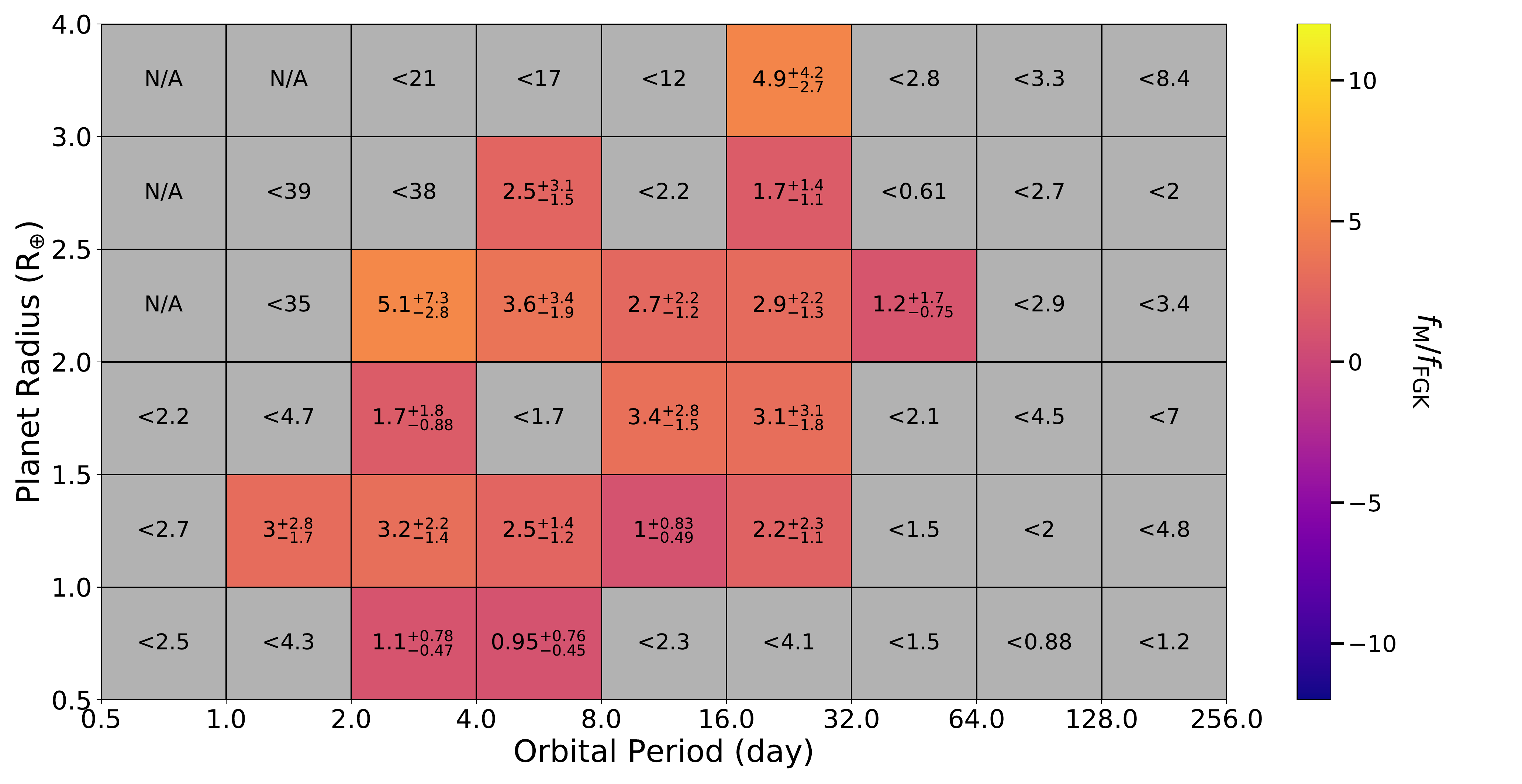}
\includegraphics[scale=0.43]{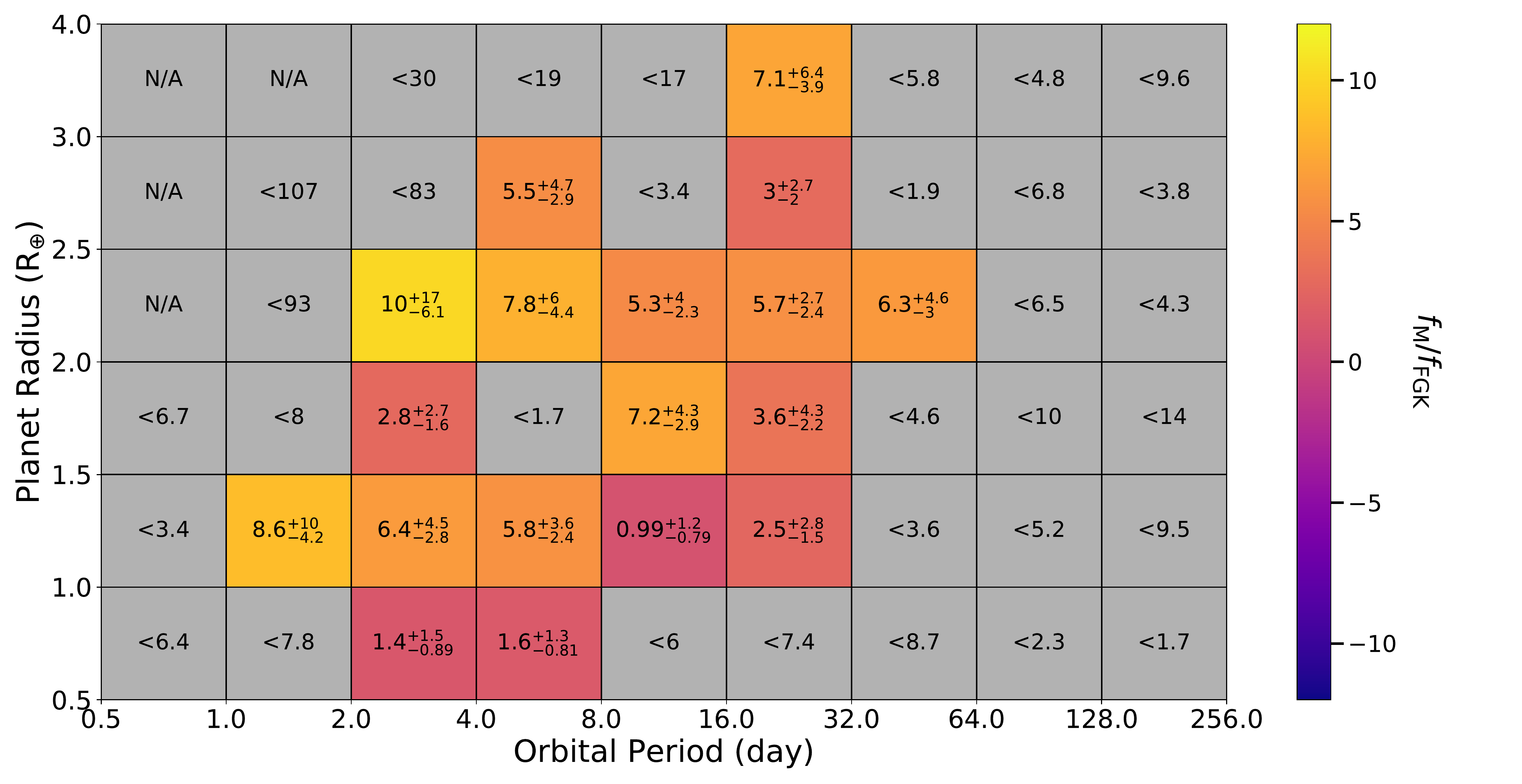}
\caption{Ratio of inferred occurrence rates for \emph{Kepler}'s DR25 planet candidates associated with high-quality M target stars to occurrence rates for DR25 planet candidates associated with high-quality FGK target stars taken from \citet{HFR+2019}.  M dwarf occurrence rates use (top) a Dirichlet prior over multiple radius bins per period range and (bottom) independent uniform priors for each bin.
The FGK rates assume independent uniform priors per period-radius bin and make use of the same combined detection and vetting efficiency model that was fit to flux-level planet injection tests for high-quality FGK target stars explained in \citet{HFR+2019}.
The average ratio across the Dirichlet prior grid is $\sim~3$ while the average ratio across the Dirichlet prior grid is
$\sim~8$,
although the scatter of ratios is large regardless of prior.
}
\label{figMFGKRatio}
\end{figure*}

In Table \ref{tab:occ_rates} we additionally report ratios of M dwarf planet occurrence rates from this study with our previously estimated planet occurrence rates for FGK stars from \citet{HFR+2019} Table 2 under the column ``Combined Detection \& Vetting Efficiency.''  These FGK planet occurrence rates are based on using the uniform prior.
Experiments showed that the uniform and Dirichlet prior results were consistent within uncertainty for all but the bins with the smallest sizes ($R_p < 1.5 \textrm{R}_\oplus$) and longest periods ($P > 128$ days). 
 
We find that using the Dirichlet prior M rates (and only including occurrence rate bins with $\geq 2$ DR25 planet candidates), the M star rate is a factor of 
$3.1^{+5.5}_{-1.9}$
larger than the associated FGK star rate for the same period and radius (as can be seen in the top panel of Fig. \ref{figMFGKRatio}). In contrast, when using the M rates estimated with the uniform prior and again filtering on bins with $\geq 2$ DR25 planet candidates we instead find the M rate is
$7.8^{+12.9}_{-5.1}$
times as large as the FGK rate over those bins (see bottom panel of Fig. \ref{figMFGKRatio}).

Another useful comparison is to contrast integrated rates calculated by summing over multiple period-radius bins. 
Throughout the remainder of the discussion, integrated rates include all bins regardless of the number of DR25 planet candidates.
Therefore, the posterior medians of upper limit bins are used in the integration and the uncertainties of each integrated rate are based on the difference of each median with the 15.87th and 84.13th percentiles from the same simulation.
Integrating over the entire period-radius grid ($P = 0.5-256$ days and $R_p = 0.5-4~\textrm{R}_\oplus$), we find
$f_{\textrm{M}} = 4.2^{+0.6}_{-0.6}$ and $8.4^{+1.2}_{-1.1}$
for planet candidates around M dwarfs with the Dirichlet and uniform priors respectively compared to the integrated rate of $f_{\textrm{FGK}} = 3.5^{+0.7}_{-0.6}$ for planet candidates around sun-like stars.
For a more robust comparison, we identify portions of the period-radius grid where most bins have $\geq 2$ DR25 planet candidates per bin (i.e., bins were we report median estimates).  Integrating over all bins with $P = 2-32$ days and $R_p = 1-2.5~\textrm{R}_\oplus$, we find 
an integrated M dwarf planet occurrence rate of
$f_{\textrm{M}} = 0.9^{+0.2}_{-0.1}$ and $1.6^{+0.3}_{-0.2}$
given a Dirichlet and uniform prior respectively, while the FGK dwarf integrated rate is $f_{\textrm{FGK}} = 0.37^{+0.04}_{-0.03}$.
Regardless of prior choice, if occurrence rates are compared in orbital period space then M dwarf occurrence rates are elevated compared to FGK dwarf occurrence rates, with the uniform prior M dwarf results suggesting a larger factor.

We also compare our estimated M rates to FGK rates by scaling the two grids to cover equivalent insolation ranges along the period axis. For this calculation, the period limits for the FGK period-radius occurrence rate grid of \citet{HFR+2019}, $P = \{2, 4, 8, 16, 32, 64, 128, 256, 500\}$ days are scaled to the insolation of a ``typical'' M dwarf in our sample\footnote{The two shortest period bins from \citet{HFR+2019} are excluded from this grid because equivalent insolation periods for those bins are $<0.5$ days, a range which our selection of window functions does not cover.}.  
We select the G2 and M2.5 spectral types as representative types for FGK dwarfs and M dwarfs respectively. The Sun (G2) is representative of sun-like stars by construction.  The distribution of $Ks$ magnitude and $J-Ks$ color for our sample indicate that the sample is primarily composed of early-type M dwarfs, and we identify M2.5 as the representative of \emph{Kepler} M dwarfs. 
The resulting period limits for a typical M2.5 target after scaling for equivalent insolation are $P = \{0.88, 1.8, 3.5, 7.1, 14, 28, 57, 113, 221\}$ days that are used for new simulations that recalculate occurrence rates over the new period-radius grid.
Here we have defined the typical FGK dwarf and typical M dwarf in our samples using \citet{M2019} values associated with the G2 ($\log L_{\textrm{FGK},\odot} = 0$, $M_\textrm{FGK} = 1~\textrm{M}_\odot$) and M2.5 ($\log L_{\textrm{M},\odot} = -1.7$, $M_\textrm{M} = 0.4~\textrm{M}_\odot$) spectral types.  
While the insolation at a given period varies between targets, we find it valuable to compare occurrence rates using these scaled period bins which are representative of the typical M dwarf target.

When comparing the M dwarf occurrence rates to FGK dwarf occurrence rates with bins chosen to have similar insolation, the ratios of M dwarf occurrence rates to associated FGK dwarf binned occurrence rates averaged over all period-radius bins with reported median rates are
$0.9^{+1.6}_{-0.2}$ and $1.7^{+1.2}_{-0.8}$
for the Dirichlet and uniform prior M dwarf estimates respectively with the uncertainties representing the standard deviation of the spread of ratios for those bins.  Integrating over the entire M dwarf period-radius grid (including upper limit bins), we find 
$f_{\textrm{M}} = 3.9^{+0.5}_{-0.5}$ and $7.0^{+0.9}_{-0.8}$
for planet candidates around M dwarfs with the Dirichlet and uniform priors respectively.   
These rates can be compared to the associated integrated rate of 
$f_{\textrm{FGK}} = 4.9^{+0.8}_{-0.7}$ 
for planet candidates around FGK dwarfs using a uniform prior.
Thus, the ratio of the integrated M dwarf occurrence rate to the integrated FGK dwarf occurrence rate is $0.8 \pm 0.2$ with the Dirichlet prior, consistent with both unity and $0.9$, the typical ratio of occurrences for individual bins with well-measured occurrence rates when using Dirichlet priors.  Using the uniform priors gives a ratio of integrated occurrence rates for M and FGK dwarfs of $1.4 \pm 0.3$, similar to $1.7$, the typical ratio of occurrences for individual bins with well-measured occurrence rates using uniform priors.
The ratios of integrated occurrence rates for M and FGK dwarfs are close to and bracket unity for the two priors.  

The host star plays an important role in the evolution of its protoplanetary disc for multiple reasons, including determining planetary compositions, affecting the location where grains can condense from the disk, the clearing of the disc via stellar winds \citep{EP2017}, and the timescale for the end of planetesimal accretion.
The close agreement between M dwarf and FGK dwarf occurrence rates found in this study when using an equivalent insolation-radius grid suggests that stellar irradiance either plays a significant role in the planet formation process or is correlated with other planet formation processes. 
We caution that planet formation occurs while an M star is still on the pre-main sequence and significantly more luminous than observed today.  
Additionally, the time of planet formation relative to the stellar evolution will depend on star and disk properties, and photoevaporation depends primarily on the XUV irradiation environment, rather than the bolometric irradiation.  
Nevertheless, we make comparisons based on the main sequence luminosity, as the luminosity at the time of planet formation is highly uncertain given the challenges of measuring ages of pre-main sequence M stars relative to the onset of planet formation.

\subsection{Occurrence Rates of Planets near the Habitable Zone}
\label{secEtaEarth}
Among M stars, the location of the habitable zone is a strong function of  each star's luminosity.  
Thus, we cannot perform the same analysis as \citet{HFR+2019} wherein we estimate the occurrence rate over a fixed radius and period range for all stars in the catalog.  
Instead, we take the estimated rates for M stars over the period-radius grid and draw 500 sample planetary systems for each star and estimate the average number of planets in the habitable zone (defined as between the runaway greenhouse and maximum greenhouse limits from \citep{KRK+2013}) on a star-to-star basis.
To report a single HZ occurrence rates, we marginalize over all stars in our target M star catalog. 
 Note that this marginalizes over occurrence rate for M dwarfs of different spectral sub-types.  Previous studies have reported significant variation in occurrence rates between early and mid M dwarfs \citep[e.g.,][]{HCM+2019}.  However, the potential impact of variation across M dwarf sub-types will likely be reduced substantially since our sample is dominated by early M dwarfs ($\sim 80\%$ of our M dwarf target stars are earlier than the M4.5 sub-type using the \citet{M2019} main sequence fit over $J-Ks$ color and $M_{Ks}$ magnitude).  

We find an estimated habitable planet occurrence rate for M stars of
$f_{\textrm{M,HZ}} = 1.1^{+0.1}_{-0.3}$ and $1.5^{+0.4}_{-0.4}$
for planet radii $R_p = 0.5-4 R_\oplus$ when simulating catalogs with Dirichlet and uniform prior estimated rates from Table \ref{tab:occ_rates} respectively.  If one places 
stricter Earth-size
constraints on planet radii requiring $R_p = 0.75-1.5 R_\oplus$, then the habitable planet occurrence rate for M stars is reduced to
$f_{\textrm{M,HZ}} = 0.33^{+0.10}_{-0.12}$ and $0.44^{+0.29}_{-0.17}$
using the Dirichlet and uniform prior rates from Table \ref{tab:occ_rates} respectively.

\subsection{Comparison to Previous Studies}
\label{secCompStudy}
For a comparison where our study can report a median estimate, we choose to compare against occurrence rate estimates from \citet{DC2015,MPA2015} over the ranges $P < 50$ days and $R_p = 1-2.5~\textrm{R}_\oplus$.  For these period-radius ranges, \citet{DC2015} finds an occurrence rate of $1.38^{+0.11}_{-0.09}$ (taken from Table 5) while \citet{MPA2015} finds a rate of $\sim 1.2\pm0.1$ (estimated from Fig. 5).  For this study, integrating over the same radii range and $P = 0.5-64$ days results in estimates of 
$1.13^{+0.20}_{-0.19}$ and $2.16^{+0.30}_{-0.34}$
respectively using the Dirichlet and uniform priors.  
Our preferred estimates based on the Dirichlet priors are consistent with the rates from \citet{DC2015,MPA2015} while the uniform prior estimate is significantly higher.  
Regardless of the best estimate, our results suggest that true uncertainties in the M dwarf occurrence rate from \emph{Kepler} are significantly larger than suggested by previous studies.

\citet{DC2015} found an integrated occurrence rate over their
period-radius grid ($P = 0.5-200$ days and $R_p = 1 - 4~\textrm{R}_\oplus$) of $2.5\pm0.2$ planets.
\citet{GMK+2016} found a rate of $f = 2.2\pm0.3$ over $R_p = 1-4~\textrm{R}_\oplus$ and $P = 1.2-180$ days, consistent with that of \citet{DC2015}.  
Over a comparable portion of the 
period-radius grid in this study ($P = 0.5-256$ days and $R_p = 1-4~\textrm{R}_\oplus$) we find occurrence rate estimates of 
$f_\textrm{M} = 2.7^{+0.4}_{-0.4}$ and $5.3^{+0.7}_{-0.7}$
using the Dirichlet and uniform priors respectively.
These estimated rates are larger than the respective estimates from \citet{DC2015,GMK+2016}, but the results using the Dirichlet prior are within $1\sigma$ agreement. 
The differences are likely explained by the differences in the target samples, the planet detection and vetting processes, and the detection efficiency models between the three studies.
We also note that our analysis results in larger uncertainties.  This is partially due to our cleaned stellar sample resulting in fewer target stars in our sample.
Another important factor is our rigorous accounting of uncertainties due to finite sample size.

Most previous studies chose to focus on occurrence rates in period-space rather than insolation-space, making comparisons over a radius-insolation grid difficult.  \citet{DC2015} is one study, however, that also infers occurrence rates over a radius-insolation grid.
However, the bin limits in insolation from \citet{DC2015} are not congruous with the period limits for equivalent insolation selected in this study.  
The closest comparisons that can be done between \citet{DC2015} and our study use the \citet{DC2015} cumulative binned rates over $10-200~F_\oplus$ in Table 7 and our binned rates over $P_\textrm{M} = 3.5-28$ days  (equivalent insolation to approximately $10.2-163~F_\oplus$ for an M2.5 dwarf).  Our study's rates are systematically larger than the \citet{DC2015} rates by factors $>2$ regardless of prior choice. 
This systematic difference likely reflects the approximation of our study to treat all stars in our M dwarf sample as having an irradiance equivalent to a \citet{M2019} M2.5 dwarf when in truth M dwarfs cover a wide range of irradiance.

Additionally, \citet{DC2015} reported an occurrence rate of $f = 0.56^{+0.06}_{-0.05}$ for Earth-size ($R_p = 1-1.5 R_\oplus$) planets with periods $<50$ days.  For our occurrence rate estimates, integrating over the same radii range and $P = 0.5-64$ days produces 
$f_\textrm{M} = 0.43^{+0.13}_{-0.10}$
using the Dirichlet prior, consistent with \citet{DC2015}.
However, we caution that using a uniform prior for each rate results in a higher estimate of 
$f_\textrm{M} = 0.74^{+0.22}_{-0.21}$.

\citet{DC2015} also gives an estimated habitable zone occurrence rate using the \citet{KRK+2013} limits that we apply in \S\ref{secEtaEarth}.  For \citet{DC2015} the habitable zone occurrence rate for Earth-size ($R_p = 1-1.5 R_\oplus$) planets was $f_{\textrm{HZ}} = 0.16^{+0.17}_{-0.07}$, whereas we find a rate of 
$f_{\textrm{M,HZ}} = 0.18^{+0.10}_{-0.05}$ or $0.17^{+0.33}_{-0.07}$
for the same radius range (for the Dirichlet and uniform priors, respectively). 
Given the associated uncertainties, the HZ occurrence rate estimates are consistent with the \citet{DC2015} estimate regardless of prior choice.

\citet{HCM+2019} finds an occurrence rate for mid-type M stars of $1.19^{+0.70}_{-0.49}$ with a limited range of $P = 0.5 - 10$ days and $R_p = 0.5 - 2.5 \textrm{R}_{\oplus}$. 
For the same radius range and $P = 0.5 - 8$ days, 
we find an inconsistent rate of
$0.47^{+0.10}_{-0.09}$
with the Dirichlet prior and a more consistent rate of 
$0.86^{+0.22}_{-0.18}$
with the uniform prior.  The latter estimate from this study is more consistent with \citet{DC2015,MPA2015}.  It is important to remember that \citet{HCM+2019} investigated mid-type M stars, so there is reason to anticipate differences in 
this
measurement.  
However, we caution that the \citet{HCM+2019} estimates are based on a small catalog of only seven stars with 13 planets, so we anticipate large uncertainties and cannot make a robust comparison of occurrence rates for these two stellar samples.

The planet radius valley \citep{FPH+2017,VAL+2017,HFR+2019} for planets around FGK dwarfs is often cited as evidence for the influence of stellar irradiance on planet formation. 
While photoevaporation \citep[e.g.,][]{OW2017,LR2018} is often cited as a potential mechanism for the planet radius valley, other physical mechanisms have been proposed including core-powered mass loss \citep{GSS2018,GS2020,GS2018}, impact erosion via planetesimals \citep{S2009,SSY2015,WKS2020}, and the formation of two distinct exoplanet populations due to a gas-poor environment for the rocky planets \citep{LCO2014,LC2016,LR2018}.
Comparing the location of the radius valley as a function of stellar type could be valuable for distinguishing between mechanisms for creating a radius valley.  
In this study, we do not find a statistically significant planet radius valley at any period.  
This contrasts with \citet{CM2019} who recently identified the planet radius valley in occurrence rates inferred from \emph{Kepler} and \emph{K2} planet candidates associated with cool stars. The slope of the valley estimated by that study supports the gas-poor formation theory explanation for the M dwarf planet radius valley, which is in contrast to the photoevaporation theory commonly invoked to explain the planet radius valley for exoplanets around FGK dwarfs.  The differences in the significance of the radius valley found in this study and \citet{CM2019} is at least partially due to the inclusion of \emph{K2} targets by \citet{CM2019}.  However, we caution that the significance of the radius valley may have been overestimated due to details of the statistical methodology in the \citet{CM2019}.  

\subsection{Future Prospects for Occurrence Rate Studies}
\label{secFutureRates}
While this work has estimated robust occurrence rates for M dwarf exoplanets, the limited number of M dwarfs in the \emph{Kepler} target catalog restricts the precision of the rates.  Future work should consider applying ABC-PMC or alternative Bayesian methodology with larger catalogs, both in number of stars and number of planet candidates.  
Recent work in formulating a pipeline for identifying and vetting of exoplanet candidates from \emph{K2} datasets 
 \citep{KMQ+2019,ZHC+2020} represents an important step in this direction.
 Creating a homogeneous exoplanet catalog from the \emph{K2} observations could provide basis for future future M dwarf occurrence rate calculations. 
In the near future, several ground-based RV (e.g. HPF, SPIRou, CARMENES) and transit surveys (e.g. MEarth, SPECULOOS), as well as the space-based \emph{TESS} mission, will increase the catalog of exoplanet candidates associated with M dwarfs. 

If the sample of exoplanets from any single catalog is too small for precise occurrence rate estimates, then another possible direction for future work is the development of a model that accounts for the unique detection pipelines for each survey.  Such a model would enable multiple catalogs as inputs to the same occurrence rate inference process, boosting the sample size and therefore increasing the precision of any estimated rates.  Previously the \citet{CG2016} study has done similar work by fitting a simple joint power-law planet distribution function to large-separation planets found by five surveys using microlensing, radial velocity, and direct imaging. A SysSim model that takes multiple catalogs would expand on \citet{CG2016} by not requiring a simple parametric planet population model and would build on \citet{CG2016} by not requiring a simple parametric planet population model and by incorporating rich information about closer-in planets from transit surveys.

The model used in this study could also be improved further to account for additional effects and reduce the number of assumptions made.  
In addition to the limitations covered in \S 4.3 of \citet{HFR+2019}, we summarize additional improvements future studies can make upon our model specifically for M star occurrence rates.

\label{secLimitations}
\label{secImprovingDetectionModel}
{\em Detection Model:}
In this study, we performed tests to verify that the detection efficiency curve fit for FGK stars was consistent with the M star sample of this study based on the available \emph{Kepler} DR25 data products.  Future research could perform more transit injection and recovery tests for M dwarf targets to enable a more detailed characterization the detection efficiency specifically for M dwarfs.  Additionally, our current detection efficiency model uses the expected effective SNR and the number of transits observed to determine the probability that a planet is detected.  As suggested in \citet{HFR+2019}, additional information (e.g., stellar properties, sky group\footnote{Sky group is an integer that groups target stars based on where they fall within the \emph{Kepler} FOV and is defined in \citet{TFV+2016}.})
could be incorporated to improve the model.

\label{secPlanetSearch}
{\em Planet Search Pipeline:}
We use planet candidates identified by the \emph{Kepler} DR25 planet search pipeline as our sample of observed planets for estimating occurrence rates.  Given that the \emph{Kepler} pipeline and robovetter were optimized to search for planets around sun-like (i.e. FGK type) targets, there may be some M dwarf associated planets that were not properly identified or vetted.  As a recent example, \citet{VRB+2020} identified a habitable zone Earth-size planet candidate associated with an M dwarf that was incorrectly classified as a false positive by the pipeline primarily due to improper photometric aperture selection.  A larger sample of identified planet candidates would improve occurrence rate precision, so future studies should consider constructing a custom planet search pipeline optimized for M dwarf targets.

\label{secFutureContam}
{\em Contamination by Non-MS M Stars:}  
Obtaining precise planet occurrence rates requires having a large sample of target stars.  
Given the \emph{Kepler} targets, one could avoid contamination of late-K stars into the M dwarf sample by applying stricter cuts to select only targets with color and absolute magnitude corresponding to spectral types later than M0. Applying such cuts based on the main sequence from \citet{M2019} would have significantly reduced the number of targets (to nearly a third of the current target list) and further reduced the precision of occurrence rate estimates.
Therefore, we chose cuts that balanced the desire for a sufficiently large target star catalog with the desire to limit the contamination of late K dwarfs. 
While we expect that the occurrence rates vary smoothly with spectral type and thus the rates for late K dwarfs should be similar to those of main-sequence M stars, we encourage future studies incorporating additional data sources to identify and analyze planet occurrence rates for target star samples that include a larger number of M dwarfs and have lower contamination from late-K dwarfs.

\label{secStellarMultiplicity}
{\em Stellar Multiplicity:}
In this study, we assume that the selected Kepler targets are single stars.  Ground-based follow-up observations can effectively rule out many types of binary stars, given the small distance to the Kepler M dwarfs.  Nevertheless, close binaries could have eluded detection and may result in spurious stellar and planetary radii.  Future studies could provide more precise constraints on a target-by-target basis and/or account for the potential effects of unrecognized binaries.

\label{secStellarNoise}
{\em Intrinsic Stellar Activity:}
Due to M stars having larger convective zones (and later types being fully convective), many M stars have significant amounts of stellar activity.   
Our detection efficiency model makes use of the one-sigma depth function for each star.  Therefore, our model already accounts for a reduction in detection and vetting efficiency due to increased stellar variability of M stars relative to FGK stars.
Nevertheless, it's possible that the rate of false positives around M stars could be increased due to the different nature of stellar activity for M dwarfs and the fact that the robovetter parameters were tuned based on FGK stars, rather than M stars mean that.  We expect these differences to be small.    
In this study we make the assumption that all planet candidates that pass the robovetter are true planets.  
A future study could perform more detailed analyses to better vet planet candidates accounting for the properties of M dwarf targets.

\label{secTidalLocking}
{\em Tidal Locking:}
The original \citet{KRK+2013} study defining HZ limits as a function of stellar temperature did not consider the effects of tidal locking.  
The habitable zone around M stars is much closer to the host star (even in units of stellar radii) than for FGK stars.
Therefore, planets at the inner edge of the \citet{KRK+2013} HZ could become tidally locked, which has implications on the habitability of the planet.  
For our study we make use of the \citet{KRK+2013} HZ limits, but a future study may consider an updated set of HZ limits that accounts for tidal locking \citep[e.g.,][]{KWH+2016}.

\subsection{Conclusions}
\label{secConclusions}
We have estimated occurrence rates for Earth to sub-Neptune sized planets over periods ranging from half a day to $\sim70\%$ a year using a cleaned sample of M dwarfs targeted by \emph{Kepler} with stellar parameters updated using \emph{Gaia} DR2 and 2MASS PSC.
These planet occurrence rates were estimated using ABC with a forward model that incorporates the final \emph{Kepler} DR25 data products to accurately reproduce the \emph{Kepler} planet candidate identification and vetting process.

We find significant differences between estimated rates using two different priors, emphasizing the importance of prior choice in the inference of occurrence rate estimates for M dwarfs in the \emph{Kepler} sample. 
Over the orbital period range of $P = 0.5-256$ days and planet radius range of $R_p = 0.5-4~\textrm{R}_\oplus$, we find an occurrence rate of 
$f_{\textrm{M}} = 4.2^{+0.6}_{-0.6}$ and $8.4^{+1.2}_{-1.1}$
when applying the Dirichlet or uniform prior respectively. 
Combining the integrated planet occurrence rates above with an the average multiplicity (i.e., planets per star with at least one planet), we can estimate the fraction of M dwarfs with planets. 
If one adopts a multiplicity of $6.1\pm1.9$ from \citet{BJ2016}, then the fraction of M dwarfs with planets (in the above range of sizes and periods) is estimated to be 
$0.7\pm0.3$ or $1.4\pm0.5$
when assuming the Dirichlet and uniform priors, respectively. 
Clearly, the fraction of stars with planets can not exceed unity.
It is important to note that the integrated planet occurrence rate includes bins with only upper limits.  In our Bayesian approach, we  marginalize over the posterior distribution of occurrence rates, extending to rates too high to be consistent with the assumed multiplicity.  
We conclude that the observed number of M dwarf planet candidates is consistent with all M dwarfs hosting a planetary system with either of our priors.
When using the Dirichlet prior,  the observed number of M dwarf planet candidates is also consistent with as few as half of M dwarfs hosting planetary systems.
Recently, \citet{HeFR2019} investigated the architectures of planetary systems around Sun-like (FGK) stars with a clustered point process model built in the same SysSim framework as this study.
\citet{HeFR2019} report a 68.3\% credible interval for the multiplicity of $2.28^{+0.94}_{0.53}$.  Since this is less than the integrated rate of M dwarf planets (for both our priors), even assigning every M dwarf such a planetary system would not be sufficient to explain the number of M dwarf planets detected by \emph{Kepler}.
We caution that a substantial fraction of the integrated M dwarf planet occurrence rate comes from bins with upper limits.  Therefore, the observations do not provide evidence the necessitates every M dwarf host a planetary system.
Collectively these three studies suggest that most M dwarf hosts at least one Earth to sub-Neptune size planet, if not several. 

We find that the planet occurrence rate around M dwarfs is substantially elevated relative to the planet occurrence rate around FGK dwarfs at similar orbital periods orbital period.  
However, the planet occurrence rates are nearly consistent (to within statistical uncertainties) when comparing at similar insolation values. 
This suggests that stellar irradiance has a significant and possibly dominant role in planet formation processes regardless of spectral type.  

Finally, we recommend that mission design concepts with the goal of characterizing M star habitable zones select science programs robust to a true rate of 
$f_{\textrm{HZ}} = 0.33^{+0.10}_{-0.12}$
for planets with $0.75-1.5$ R$_\oplus$ size.

\section*{Acknowledgments}
We thank the entire \emph{Kepler} team for the many years of work that has proven so successful and was critical to this study.
We acknowledge valuable feedback from Suvrath Mahadevan, Darin Ragozzine and Jason Wright.  
D.C.H and E.B.F. acknowledge support from NASA Origins of Solar Systems grant \# NNX14AI76G and Exoplanet Research Program grant \# NNX15AE21G.
D.C.H and E.B.F. acknowledge support from the Penn State Eberly College of Science and Department of Astronomy \& Astrophysics, the Center for Exoplanets and Habitable Worlds and the Center for Astrostatistics.  
The citations in this paper have made use of NASA's Astrophysics Data System Bibliographic Services.  
This research has made use of the NASA Exoplanet Archive, which is operated by the California Institute of Technology, under contract with the National Aeronautics and Space Administration under the Exoplanet Exploration Program.
This work made use of the gaia-kepler.fun crossmatch database created by Megan Bedell.
We acknowledge the Institute for Computational and Data Sciences (\url{http://ics.psu.edu/}) at The Pennsylvania State University for providing advanced computing resources and services that have contributed to the research results reported in this paper.
This study benefited from the 2013 Statistical and Applied Mathematical Sciences Institute (SAMSI) workshop on Modern Statistical and Computational Methods for Analysis of \emph{Kepler} Data, the 2016/2017 Program on Statistical, Mathematical and Computational Methods for Astronomy, and their associated working groups.  

\bibliographystyle{mnras}
\bibliography{references}

\bsp	
\label{lastpage}
\end{document}